\documentclass[a4paper]{article}
\usepackage{jheppub}
\usepackage{amssymb}   
\usepackage{amsmath}
\usepackage{mathtools}
\usepackage{makeidx}
\usepackage{graphicx}
\usepackage{caption}
\usepackage{subcaption}
\usepackage[utf8]{inputenc}
\usepackage{feynmp-auto}
\usepackage{graphicx}
\usepackage[draft]{fixme}
\usepackage{xcolor}

\def\beq{\begin{equation}}
\def\eeq{\end{equation}}
\def\bea{\begin{eqnarray}}
\def\eea{\end{eqnarray}}

\def\f21{{}_2F_{1}}

\def\O{\mathcal{O}}

\hypersetup{colorlinks = false}

\def\bsp#1\esp{\begin{split}#1\end{split}}
\newcommand{\zb}{\bar{z}}

\preprint{CERN-TH-2017-200, SLAC-PUB-17158}

\title{Differential Higgs production at N$^3$LO beyond threshold}

\author[a]{Falko Dulat}
\author[b]{Bernhard Mistlberger}
\author[c]{Andrea Pelloni}
\affiliation[a]{SLAC National Accelerator Laboratory, Stanford University, Stanford, CA 94039, USA}
\affiliation[b]{CERN Theory Division, CH-1211, Geneva 23, Switzerland}
\affiliation[c]{Institute for Theoretical Physics, ETH Z\"urich, 8093 Z\"urich, Switzerland}
\emailAdd{dulatf@slac.stanford.edu}
\emailAdd{bernhard.mistlberger@cern.ch}
\emailAdd{apelloni@student.ethz.ch}

\abstract{
We present several key steps towards the computation of differential Higgs boson cross sections at N$^3$LO in perturbative QCD.
Specifically, we work in the framework of Higgs-differential cross sections that allows to compute precise predictions for realistic LHC observables.
We demonstrate how to perform an expansion of the analytic N$^3$LO coefficient functions around the production threshold of the Higgs boson.
Our framework allows us to compute to arbitrarily high order in the threshold expansion and we explicitly obtain the first two expansion coefficients in analytic form.
Furthermore, we assess the phenomenological viability of threshold expansions for differential
distributions.
We find that while a few terms in the threshold expansion are sufficient to approximate the
    exact rapidity distribution well, transverse momentum distributions require a signficantly
    higher number of terms in the expansion to be adequately described. 
    We find that to improve state of the art predictions for the rapidity distribution beyond NNLO even more sub-leading terms in the threshold expansion than presented in this article are required.
In addition, we report on an interesting obstacle for the computation of N$^3$LO corrections with LHAPDF parton distribution functions and our solution.
We provide files containing the analytic expressions for the partonic cross sections together with the arXiv submission.
}


\begin{document}
\notoc
\maketitle
\section{Introduction}
The discovery of the Higgs boson in 2012 by the ATLAS~\cite{Aad2012} and CMS~\cite{Chatrchyan2012} experiments at the Large Hadron Collider (LHC) was a major scientific breakthrough.
Confirming for the first time conclusively the presence of the Higgs field and explaining the origin of the masses of elementary particles, it renders the standard model mathematically self-consistent, allowing it to be used to formulate credible predictions at high energies.
At the same time it is clear that the standard model needs to be extended to explain the puzzle of neutrino masses and cosmological observables as well as the nature of dark matter.

If these open questions are related to the origin of masses it is likely that the precise nature of the Higgs boson will differ quantitatively from what is expected in the standard model. Lacking direct observations of particles beyond the standard model, the Higgs boson is the most promising probe into possible ultra-violet completions of the standard model.
As such, a precise study of the properties of the Higgs boson, such as mass, spin/parity, branching ratios and production rates, is a central part of the present and future precision Higgs physics program at the LHC that has been initiated since the discovery.
The remarkable performance of the LHC, delivering unprecedented luminosity, as well as the unrelenting efforts of the experiments have led to vast improvements of Higgs analyses resulting in more and more precise extractions of the properties of the Higgs boson.

In order to truly exploit the potential of these excellent experimental results, it is imperative to confront them with equally precise theoretical predictions.
This demand for precise predictions from theory has lead to a flurry of calculations in recent years at next-to-leading (NLO) and next-to-next-to-leading (NNLO) order in perturbative QCD. Higher order predictions are particularly important for Higgs phenomenology, in part due to the slow convergence of the perturbative expansion in the strong coupling constant for the dominant mode of Higgs hadroproduction via gluon fusion.
The phenomenological importance of these corrections can be seen from the large size of the NLO corrections~\cite{Dawson:1990zj,Spira:1995rr} which almost double the leading-order prediction~\cite{Wilczek1977}. The magnitude of these perturbative corrections indicate potentially significant contributions from even higher orders in the perturbative series, leading to a substantial uncertainty on the gluon-fusion cross section. Even with the inclusion of the NNLO~\cite{Harlander:2002wh,Anastasiou2002,Ravindran:2003um} corrections to the gluon-fusion cross section the perturbative series seems to converge slowly keeping the perturbative uncertainty at a significant level.

Recently, the gluon-fusion cross section has been computed through next-to-next-to-next-to-leading order (N$^3$LO) in perturbative QCD~
\cite{Anastasiou:2014vaa, Anastasiou:2014lda, Anastasiou:2015ema}
in the limit of an infinite top mass. This calculation has lead to a reduction of the perturbative uncertainty, making the theoretical predictions competitive with current experimental analyses. At this level of precision, effects that go beyond the leading approximation of an infinite top mass or a treatment in pure QCD, neglecting effects from quark masses or electroweak loops, become important. The state of the art predictions for Higgs hadroproduction have been combined in a consistent way in~\cite{Anastasiou:2016cez} and are also compiled in~\cite{deFlorian:2016spz,Harlander:2016hcx}.

The Higgs cross section is measured in dedicated regions of phase space as required by the detector geometry and optimized by carefully designed experimental event selections. Within these acceptance regions, the experiments have excellent capabilities to measure a plethora of kinematic distributions for the Higgs boson and its decay products that can be used to characterize the properties of the Higgs boson.
As such it is imperative for theoretical predictions to not only inclusively describe the total cross section but also provide high precision theoretical predictions for differential cross sections.
Recently, the $pp\to H+1\textrm{ jet}$ fully differential cross section has been computed at NNLO~\cite{Chen2015,Boughezal2015b,Boughezal2015a}. In combination with the inclusive N$^3$LO cross section this enables the computation of the jet-vetoed Higgs cross section at N$^3$LO~\cite{Banfi2016}. Fully differential parton-level Monte-Carlo simulations at N$^3$LO will enable the study of efficiencies of many other event selection criteria at the same accuracy in perturbation theory as the jet-veto efficiency.

One way to achieve this goal would be to attempt to generalize any of the methods available at NNLO  (sector-decomposition~\cite{Anastasiou:2005cb,Binoth:2000ps,Hepp:1966eg,Roth:1996pd},
slicing~\cite{Boughezal2011,Boughezal2015,Catani2007,Gaunt2015}, subtraction~\cite{Baernreuther2012,Caola2017,i2007,Ridder2005},
reweighting~\cite{Cacciari:2015jma} and other methods~\cite{Anastasiou2010}) to N$^3$LO. However, the adaptation of any of these methods to N$^3$LO in full generality is a formidable task. 
Another avenue is to focus on some specific differential observables at N$^3$LO. 
To this end we introduced the \emph{Higgs-differential} method~\cite{Dulat:2017aa} and studied its feasibility in obtaining differential distributions without a sophisticated NNLO subtraction method. 
The method is fully-differential in all components of the Higgs momentum and its decay products, extending ideas first used to obtain rapidity distributions at NNLO~\cite{Anastasiou:2002qz,Anastasiou:2003yy}.
At the same time, it treats the additional QCD radiation inclusively (i.e. it integrates over the unrestricted phase space of all final state partons). 
This enables us to compute differential Higgs boson observables, i.e. the Higgs boson rapidity and transverse momentum distributions, as well as any distributions of decay products.
At a later stage the results obtained with this method may serve as a key ingredient to a fully differential computation, for example in combination with $q_T$-subtraction~\cite{Catani2007}.

A complete analytical computation of N$^3$LO Higgs differential cross sections presents a formidable challenge.
Key ingredients for such a computation are analytic expressions for Feynman integrals in kinematic limits that can serve as boundary data as well as counter terms that render the Higgs differential cross section finite in all its physical limits.

In this article we achieve a significant step towards this computation by performing an expansion of the complete partonic cross sections around the production threshold of the Higgs boson. 
The analytic information we obtain is comprised of the first two terms of in the threshold expansion and represents the foundation of a complete future computation. 
Furthermore, we explore the phenomenological potential of threshold expansions to approximate differential Higgs boson observables. 
Computations at very high orders stress the reliability of conventional tools for higher order computations to their limits. 
We identify a pitfall with the standard treatment of parton distribution functions within the framework of LHAPDF and present our solution.
Furthermore, we report on the complete computation of contributions to the Higgs differential N$^3$LO corrections involving explicit logarithms of the perturbative scale.

The threshold prediction for the Higgs boson rapidity distribution at N$^3$LO was also obtained
in~\cite{Ravindran:2006bu,Ahmed:2014uya, Banerjee:2017cfc} in the double threshold expansion.
The logarithmic contributions to the tranverse momentum spectrum for small
transverse momenta were also obtained in~\cite{Li:2016axz,Li:2016ctv} within the framework of SCET.
We would like to point out that our method does not rely on the simplification of the definition of the physical observables in any kinematic limit.
This allows us at least in principle to compute the Higgs differential N$^3$LO cross sections to any order in the threshold expansion.
The results that we obtain in this note are primarily a proof of principle, demonstrating
that it is possible, to extend our Higgs differential method to N$^3$LO. Consequently, we do not
claim any phenomenological significance of the distributions presented in the following. This is
applies in particular to the transverse momentum distributions that we show in
section~\ref{sec:threshold}. It is of course a well known fact that transverse momentum
distributions for Higgs production in gluon fusion in the effective field theory have a fairly
limited range of applicability. In the low end of the transverse momentum spectrum, the cross section is dominated
by large logarithms of the transverse momentum that need to be resummed to all orders in
perturbation theory to obtain stable predictions. 
At the high tail of the transverse momentum
spectrum the cross section becomes sensitive to finite quark mass effects and the accuracy of the
infinity top mass approximation decreases rapidly. This can be countered by systematically adding 
corrections to the infinite top-mass limit to the calculations.
A detailed discussion of both of these issues is premature at this point, as our first goal is to
obtain a fixed order prediction for the Higgs differential distributions in the infinite top-mass
limit through N$^3$LO in QCD. Once this is achieved, it will be a separate issue to systematically
improve upon this result and combine it with known important effects as the ones mentioned above.

This article is organized as follows. In section~\ref{sec:HDiff}, we review our method of \emph{Higgs-differential} calculations. 
In section~\ref{sec:N3LOThresholdCalc} we describe how threshold expansions can be performed for partonic Higgs-differential cross sections and obtain analytic results for the N$^3$LO coefficient function.
In section~\ref{sec:threshold} we critically asses the quality of results obtained with a threshold expansion for differential distributions at NNLO.
On the way to obtain numerical results for distributions at N$^3$LO we encounter in section~\ref{sec:PDFGames} an issue that arises when interpolating parton density functions. Naive usage of LHAPDF leads to a drop in accuracy when computing N$^3$LO results due to a lack of smoothness in the default interpolation routines. We illustrate our findings and discuss our way of avoiding this issue.
Yet another crucial ingredient for N$^3$LO Higgs-differential phenomenology are the complete contributions due to explicit dependence on the perturbative scale which we obtain in section~\ref{sec:scalevar}.
Then, in section~\ref{sec:N3LONumerics} we demonstrate the impact of the newly obtained approximations to the N$^3$LO coefficient functions on the rapidity distribution of the Higgs boson and discuss their validity.
Finally, we conclude in section~\ref{sec:conclusions}.

\newpage
\section{Higgs Differential Cross Sections}
\label{sec:HDiff}
In this section we briefly review the definition of Higgs differential cross sections introduced in ref.~\cite{Dulat:2017aa}.
Within this framework we consider scattering process of two protons that produce at least a Higgs boson.
\begin{equation}
{\rm Proton}(P_1) + {\rm Proton}(P_2) \to H(p_h) + X,
\end{equation}
$P_1$ and $P_2$ are the momenta of the colliding protons and $p_h$ the momentum of the Higgs boson.
The framework of Higgs differential cross sections allows to compute the scattering probability for any observable that is solely dependent on the four momentum of the Higgs boson.
Typically such observables are related to the rapidity $Y$, transverse momentum $p_T$ and mass $m_h$ of the Higgs boson. 
\begin{equation}
p_h \equiv \left( E,\, p_x,\, p_y,\, p_z\right) =
\left(\sqrt{p_T^2 + m_h^2} \cosh Y,\; p_T \cos\phi,\; p_T \sin\phi,\; \sqrt{p_T^2 + m_h^2} \sinh Y\right),
\end{equation}
where
\[\label{eq:hadronic_rapidity}
Y=\frac{1}{2} \log \left( \frac{E+p_z}{E-p_z} \right),
\qquad
p_T=\sqrt{E^2-p_z^2-m_h^2}.
\]
$E$, $p_z$ and $p_T$ are the energy of the Higgs boson, its momentum along  and transverse to the beam axis in the laboratory rest frame, respectively.
The master formula for a Higgs differential cross section for  an observable $\O$  is then given by
\begin{multline}
\label{eq:xsdiffhad2}
\sigma_{PP\rightarrow H+X}\left[\O\right]=
\tau \sum_{i,j} \int_\tau^1 \frac{dz}{z}\int_{\frac{\tau}{z}}^1 \frac{dx_1}{x_1} \int_0^1 dx \int_0^1 d\lambda \int_0^{2\pi}\frac{d\phi}{2\pi}\\
\times f_i(x_1)f_j\left(\frac{\tau}{x_1 z}\right)
\frac{1}{z}\frac{d^2 \sigma_{ij}}{dx\,d\lambda}(z,x,\lambda,m_h^2)
\mathcal{J}_\O(x_1,z,x,\lambda,\phi,m_h^2).
\end{multline}
Here, we employed the parton model and factorization of long and short range interactions into parton distribution functions $f_i(x)$ and partonic differential cross sections. The momenta of the colliding partons are related to the proton momenta by $p_1=x_1 P_1$ and $p_2=x_2 P_2=\frac{\tau}{x_1 z}P_2$. The variable $\tau$ is given by 
\beq
\tau=\frac{m_h^2}{S},\hspace{1cm} S=(P_1+P_2)^2.
\eeq
The sum over $i$ and $j$ ranges over all contributing partons. The variables $x$, $\lambda$ and $\phi$ parametrize the momentum of the Higgs boson. $\phi$ is the azimuthal angle of the Higgs boson with respect to the collision axis. 
$x$ and $\lambda$ are related to the more familiar Higgs boson rapidity and transverse momentum by
\begin{equation}
\label{eq:vardefpty}
Y=\frac{1}{2} \log \left[
\frac{x_1}{x_2}
\frac{1-\frac{\bar z \bar \lambda}{1-\bar z \lambda x}}{1-\bar z \lambda}
\right],
\qquad
p_T^2 = s \frac{\bar z^2  \lambda  \bar \lambda \bar x}{1-\bar z x \lambda}.
\end{equation}
Here, $\bar x=1-x$, $\bar \lambda=1-\lambda$ and $\bar z=1-z$.
The partonic Higgs differential cross section is given by $\frac{d^2 \sigma_{ij}}{dx\,d\lambda}(z,x,\lambda,m_h^2)$. The observable we are interested in is specified by the measurement function $\mathcal{J}_\O(x_1,z,x,\lambda,\phi,m_h^2)$, that filters the regions we are interested in.
Assume we are interested in computing the probability for a Higgs boson to be produced in the rapidity interval $Y\in [1,2]$
the measurement function would take the form 
\beq
\mathcal{J}_\O(x_1,z,x,\lambda,\phi,m_h^2)=\theta\left(2-Y(x_1,z,x,\lambda,\phi,m_h^2)\right)\theta\left(Y(x_1,z,x,\lambda,\phi,m_h^2)-1\right).
\eeq

In ref.~\cite{Dulat:2017aa} the partonic Higgs differential cross sections were computed in heavy quark effective theory for the gluon fusion production mode to NNLO in QCD perturbation theory in terms of analytic functions of the variables $z$, $x$ and $\lambda$.  
Higgs differential cross sections can be easily combined with subsequent decays of the Higgs boson in order to allow for the prediction of fiducial cross sections for Higgs boson decay products as demonstrated in ref.~\cite{Dulat:2017aa}.

\section{Threshold Expansion for Higgs-differential N$^3$LO}

\label{sec:N3LOThresholdCalc}
In this section we present the analytic computation for the first and second term of the threshold expansion of the partonic N$^3$LO coefficient functions. 
In order to derive our results we strongly rely on techniques recently developed in refs.~\cite{Anastasiou:2013srw,Anastasiou:2013mca,Anastasiou:2015ema,Anastasiou:2015yha,Anastasiou:2013mca}.
We begin by clarifying the ingredients for our partonic coefficient functions. 
Next, we set-up the notation for the partonic phase space integrals we perform to obtain Higgs differential partonic cross sections.
Finally, we explain how threshold expansion for Higgs differential cross sections can be performed at the integrand level.
\subsection{Setup of the Calculation}
The gluon fusion cross section in the standard model is mainly mediated by a top quark loop, coupling the gluons to the Higgs boson. For the production of a Higgs boson with the observed mass of $m_H=125\textrm{GeV}$ with a transverse momentum below the top-pair threshold, the process can be safely described in the limit where the top quark is infinitely heavy. In this limit, the top quark can be integrated out, which induces an effective theory that directly couples the gluons to the Higgs boson via an effective dimension five operator. The Lagrangian for this effective theory is given by,
\beq
\mathcal{L}_{\textrm{eff}} = \mathcal{L}_{\textrm{SM},5} - \frac{1}{4}C^0HG^{\mu\nu}_aG_{\mu\nu a},
\eeq
where $H$ is the Higgs field, $G^{\mu\nu}_a$ is the gluon field strength tensor and $\mathcal{L}_{\textrm{SM},5}$ denotes the standard model Lagrangian with $N_f=5$ light flavors.
The bare Wilson coefficient $C^0$ is obtained by matching the effective theory to the standard model in the limit of an infinitely heavy top quark~\cite{Chetyrkin:1997un,Schroder:2005hy,Chetyrkin:2005ia,Kramer:1996iq}.
The inclusive cross section $\sigma_{PP\to H+X}$ has been computed at NLO~\cite{Dawson:1990zj,Graudenz:1992pv,Spira:1995rr} as well as at NNLO~\cite{Anastasiou2002,Harlander:2002wh,Ravindran:2003um}. Recently, the N$^3$LO corrections were also computed~\cite{Anastasiou:2015ema,Anastasiou:2016cez}.

Within the effective theory, we can write the \emph{Higgs-differential} partonic cross section as,
\bea 
\frac{1}{z}\frac{d^2 \hat{\sigma}_{ij}}{dx\,d\lambda}(z,x,\lambda,m_h^2)&=& (C^0)^2 \,\hat{\sigma}_0 \, \eta_{ij}(z,x,\lambda)\nonumber\\
&=& (C^0)^2 \,\hat{\sigma}_0 \, \sum\limits_{k=0}^\infty \left(\frac{\alpha_S}{\pi}\right)^k \eta^{(k)}_{ij}(z,x,\lambda).
\eea
Dividing out the Born cross section,
\beq
\hat{\sigma}_0=\frac{\pi}{8(n_c^2-1)},
\eeq
we can write the partonic coefficient functions as,
\bea
 \eta^{(n)}_{ij}(z,x,\lambda)&=&\frac{N_{ij}}{2 m_h^2 \hat\sigma_0}\sum\limits_{m=0}^n \int d\Phi_{H+m}\delta\left(x - \frac{s(p_1+p_2-p_h)^2}{(s-2p_1 \cdot p_h)(s-2p_2 \cdot p_h)}\right) \nonumber\\
&\times&\delta\left(\lambda - \frac{s-2p_1\cdot p_h}{s-m_h^2}\right) \mathcal{M}^{(n)}_{ij\rightarrow H+m}.
\eea
The initial state dependent prefactors $N_{ij}$ are given by
\begin{align}
N_{gg}&=\frac{1}{4(1-\epsilon)^2(n_c^2-1)^2},\nonumber\\
N_{gq}&=N_{qg}=\frac{1}{4(1-\epsilon)(n_c^2-1)n_c},\\
N_{q\bar q}&=N_{qq}=N_{qq^\prime}=\frac{1}{4n_c^2}.\nonumber
\end{align}
Here, $g$, $q$ and $\bar q$ indicate that the initial state parton is a gluon, quark or anti-quark respectively.
$d\Phi_{H+m} $ is the phase space measure for the production of a Higgs boson and $m$ partons and is explained in more detail below.
 $ \mathcal{M}^{(n)}_{ij\rightarrow H+M}$ is the coefficient of $\alpha_S^n$ in the coupling constant expansion
of the modulus squared of all amplitudes for partons $i$ and $j$ producing a final state Higgs boson and $m$ partons summed over polarizations and colors. To compute the n$^{\text{th}}$ order partonic coefficient functions we require all combinations $l$-loop matrix elements with $m$ external particles such that $m+l=n$. 

The parameter $\bar z=1-z=1-\frac{m_h^2}{s}$ tends to zero as we approach the production threshold and the partonic center of mass energy becomes equal to the Higgs boson mass. 
In this work we perform a systematic expansion of the partonic coefficient functions around the production threshold.
\beq
\label{eq:etas}
 \eta^{(n)}_{ij}(z,x,\lambda)= \eta^{(n,SV)}_{ij}(z,x,\lambda)+\sum\limits_{k=0}^\infty \bar z ^i  \eta^{(n,k)}_{ij}(z,x,\lambda).
\eeq
We separated the leading term in the expansion that is indicated by the superscript $(SV)$ and that is commonly referred to as the soft-virtual contribution. 
This particular term is singular as $\bar z \rightarrow 0$ and acts as a distribution on the measurement function and the parton distribution functions as we explain below. 
All higher power terms depend on $z$ in the form of polynomials of logarithms $\log(1-z)$.
The individual terms $\eta^{(n,i)}_{ij}(z,x,\lambda)$ depend on the threshold variable $\bar z$ in the form of polynomials of logarithms of  the form $\log(\bar z)$. 
In this article we obtain the first and second term in the threshold expansion for all required partonic coefficient functions.

The purely virtual matrix elements are independent of the expansion parameter and  were computed in refs.~\cite{Gehrmann2010}. 
Matrix elements with two loop and one emission were computed in refs.~\cite{Anastasiou:2013mca,Duhr:2013msa,Duhr:2014nda} and recomputed for the purpose of this article and are known to all orders in $\bar z$. 

In order to obtain the required matrix elements with two or three additional partons in the final state we followed the techniques developed in refs.~\cite{Anastasiou:2013mca,Anastasiou:2015ema,Anastasiou:2015yha,Anastasiou:2013mca}. 
We generate Feynman diagrams with QGRAF~\cite{Nogueira1993} and perform spinor and color algebra with a private c++ code based on GiNaC~\cite{Bauer2000}.
Next, we perform a threshold expansion on the integrand level and subsequently integrate out the momenta of all radiation produced in addition to the Higgs boson.
We discuss this step in greater detail below.

Our coefficient functions contain single poles in the variables $\zb$, $x$ and $\lambda$. 
These poles correspond to kinematic singularities of the Higgs boson cross section where the kinematic degrees of freedom degenerate. Explicitly, they correspond to vanishing transverse momentum of the Higgs boson or vanishing virtuality of the system of all radiation produced in association with the Higgs boson. 
Specifically, before expansion in the dimensional regulator these singularities are of the form
\beq
\left\{\zb^{-1+a_1 \epsilon},x^{-1+a_2 \epsilon},(1-x)^{-1+a_3 \epsilon},\lambda^{-1+a_4 \epsilon},(1-\lambda)^{-1+a_5 \epsilon}\right\}
\eeq
where the coefficients $a_i$ are small integer numbers. 

When we compute Higgs differential cross sections as in eq.~\eqref{eq:xsdiffhad2} we integrate in the variables $\zb$, $\lambda$ and $x$ and the singularities may lie within our integration range, depending on the observable under consideration.
For example, to compute the inclusive cross section we integrate the variables $x$ and $\lambda$ within the interval $[0,1]$.
Consequently, regularization of these divergences is required and we proceed as outlined in ref.~\cite{Dulat:2017aa} (and repeated in appendix~\ref{sec:APPReg}). The analytic computation of the first two terms in the threshold expansion of the N$^3$LO coefficient functions represents one of the main results of this article. We present this result in Mathematica readable form as an ancillary file submitted together with the arXiv version of this article.

\subsection{Higgs Differential Phase Space}
The integration measure for the production phase space of a Higgs boson and $m$ additional partons is given by
\beq
\label{eq:measure}
d\Phi_m= \frac{d^dp_h}{(2\pi)^d} (2\pi)\delta_+(p_h^2-m_h^2)(2\pi)^d \delta^d\left(p_1+p_2+p_h+\sum\limits_{i=3}^{m+2} p_i\right)\prod\limits_{i=3}^{m+2}\frac{d^dp_i}{(2\pi)^d} (2\pi)\delta_+(p_i^2),
\eeq
where 
\beq
\delta_+(p^2-m^2)=\theta(-p^0+m)\delta(p^2-m^2).
\eeq
We want to develop a parametrization of the above phase space that allows us to compute observables that are differential in the Higgs boson four momentum. Consequently, it seems natural to separate the integration over the momenta of the Higgs and the final state parton momenta. We can achieve this by inserting a unity
\beq
1=\int \frac{d^dk}{(2\pi)^d} (2\pi)^d\delta(k-p_3-p_4-\dots-p_{m+2})\int_0^\infty \frac{d\mu^2}{2\pi}(2\pi)\delta_+(k^2-\mu^2).
\eeq
This identity allows us to write the $H$ plus $m$ parton phase space measure as an integral over a phase space measure for two massive particles and $m$ massless partons.
\beq
\label{eq:factored}
d\Phi_m=\int_0^{\infty} \frac{d\mu^2}{2\pi}d\Phi^{2-\text{m}} d\Phi^{0-\text{m}}_m.
\eeq

We choose the rest frame of the initial state partons
\beq
\label{eq:comframe}
p_1^\mu=\frac{\sqrt{s}}{2}\left(\begin{array}{c} 1 \\ 0 \\ \vdots \\ 0 \\ -1\end{array}\right), \hspace{1cm} p_2^\mu=\frac{\sqrt{s}}{2}\left(\begin{array}{c} 1 \\ 0 \\ \vdots \\ 0 \\ 1\end{array}\right).
\eeq
We introduce the following definition of Lorentz invariant scalar products.
\bea
s_{ij}&=&(p_i+p_j)^2,\hspace{1cm}i\neq j.\nonumber\\
s_{ii}&=&p_i^2.
\eea
With this we have $s=s_{12}$.
We refer to the last component of our $d$-dimensional vectors as $z$ components and all other space like components as transverse components. The z-axis is parallel to the collision axis of the incoming partons.
In this frame we can now express the energy- and z-component of any vector $p_i$ in terms of scalar products of this vector and the two incoming momenta.
\bea
p_i^0&=&\frac{1}{\sqrt{s}} (p_1+p_2)\cdot p_i.\nonumber\\
p_i^z&=&\frac{1}{\sqrt{s}} (p_1-p_2)\cdot p_i.
\eea

We start by parametrizing the phase space for two massive particles.
\bea
d\Phi^{2-m}&=&\frac{d^d p_h}{(2\pi)^d}\frac{d^d k}{(2\pi)^d} (2\pi)\delta_+(k^2-\mu^2)\,(2\pi)\delta_+(p_h^2-m_h^2)\,(2\pi)^d\delta^d(p_1+p_2+p_h+k).\nonumber\\
\eea
In general we can write 
\bea
\label{eq:1loopparam}
\frac{d^d p_i}{(2\pi)^d}&=&\frac{1}{(2\pi)^d}dE_i dp_i^z d^{d-2} p_i^{\perp} =\frac{1}{2 (2\pi)^d}dE_i dp_i^z d|p_i^{\perp}|^2 d\Omega_{d-2} \left(|p_i^\perp|^2\right)^{\frac{d}{2}-2}.
\eea
Integrating out $p_h$ and using the above parametrization we find
\bea
d\Phi^{2-m}&=&\frac{(2\pi)^{2-d}}{2} dE_k dk^z d|k^\perp |^2 d\Omega_{d-2} \left[(k^\perp)^2\right]^{\frac{d}{2}-2} \nonumber\\
&\times&\theta(-E_k-\mu)\theta(E_k+\sqrt{s}-m_h)\delta(k^2-\mu^2)\delta(s+2p_{12}k+k^2-m_h^2).
\eea
We exploit the on-shell condition of $k$ to perform the $k^\perp$ integration to find  $|k^\perp|^2=E_k^2-(k^z)^2-\mu^2$. Furthermore, we parametrize 
\bea
2p_1 k&=&-(s-m_h^2) \lambda=\sqrt{s}(E_k-k^z).\nonumber\\
2p_2 k&=&-(s-m_h^2) \bar\lambda=\sqrt{s}(E_k+k^z).
\eea
and find that
\beq
|k^\perp|^2=E_k^2-(k^z)^2-\mu^2=\frac{(s-m_h^2)^2}{s} \lambda \bar \lambda -\mu^2.
\eeq
Consequently,
\bea
d\Phi^{2-m}&=&\frac{(2\pi)^{2-d}}{4} s\bar z^2 d\lambda d\bar \lambda d\Omega_{d-2} \left[\frac{(s-m_h^2)^2}{s} \lambda \bar \lambda -\mu^2\right]^{\frac{d}{2}-2} \nonumber\\
&\times&\theta(-E_k-\mu)\theta(E_k+\sqrt{s}-m_h) \delta((s-m_h^2)(1-\lambda-\bar\lambda)+\mu^2).
\eea
In order to make our live a little easier we are going to exploit the fact that we will later on integrate over $\mu^2$ and re-parametrize $\mu^2=s \bar z^2 \lambda \bar \lambda x$:
\bea
d\Phi^{2-m}&=&\frac{(2\pi)^{2-d}}{4}s\bar z^2 d\lambda d\bar \lambda d\Omega_{d-2} \left[s \bar z ^2 \lambda \bar \lambda (1-x)\right]^{\frac{d}{2}-2} \nonumber\\
&\times&\theta(-E_k-\mu)\theta(E_k+\sqrt{s}-m_h) \delta(s \bar z (1-\lambda -\bar \lambda(1-\bar z\lambda x))).
\eea
Solving the $\theta$ constraints and integrating out $\bar \lambda $ we find 
\bea
d\Phi^{2-m}&=&\frac{(2\pi)^{2-d}}{4} s^{\frac{d}{2}-2} \bar z^{d-3} d\lambda d\Omega_{d-2}\left(\lambda (1-\lambda)\right)^{\frac{d}{2}-2}\left(1-x \right)^{\frac{d}{2}-2}\left(1-\bar z \lambda x\right)^{1-\frac{d}{2}}\nonumber\\
&\times&\theta(\lambda)\theta(1-\lambda)\theta(1-x)\theta(\bar z)\theta(s).
\eea
We now can combine the previous result and rewrite eq.~\eqref{eq:factored} as
\bea
d\Phi_m&=&\int_0^{\infty} \frac{d\mu^2}{2\pi}d\Phi^{2-\text{m}} d\Phi^{0-\text{m}}_m\nonumber\\
&=& \frac{s^{\frac{d}{2}-1}}{4(2\pi)^{d-1}}  \bar z^{d-1}d\lambda d\Omega_{d-2}\, \theta(\bar z)\theta(s) \int dx\,\theta(x)\theta(1-x)\theta(\lambda)\theta(1-\lambda)  \nonumber\\
&\times&\left(\lambda (1-\lambda)\right)^{\frac{d}{2}-1}\left(1-x \right)^{\frac{d}{2}-2}\left(1-\bar z \lambda x\right)^{-\frac{d}{2}} d\Phi^{0-\text{m}}_m.
\eea
The remaining massless parton measure is given by
\beq
\label{eq:masslessmeasure}
d\Phi^{0-m}=(2\pi)^d \delta^d\left(k-\sum\limits_{i=3}^{m+2} p_i\right) \prod\limits_{i=3}^{m+2}\frac{d^dp_i}{(2\pi)^d} (2\pi)\delta_+(p_i^2).
\eeq
Analytic partonic coefficient functions for Higgs differential cross sections are now obtained by performing the integration over $m$-parton final state squared matrix elements using the above measure.

\subsection{Threshold Expansions for Higgs Differential Cross Sections}
In this section we describe our method to perform a threshold expansion at the integrand level for the required matrix elements for the N$^3$LO coefficient function that involve two or more partons. We start by regarding matrix elements that correspond to purely real radiation diagrams and contain no closed loops and then we address the case of virtual radiations.

Consider as an example the following scalar phase space integral.
\bea\label{RRRex}
I(p_1,p_2,k)&=&\begin{gathered}\vspace{-1em}\scalebox{0.5}{\includegraphics[width=0.95\textwidth]{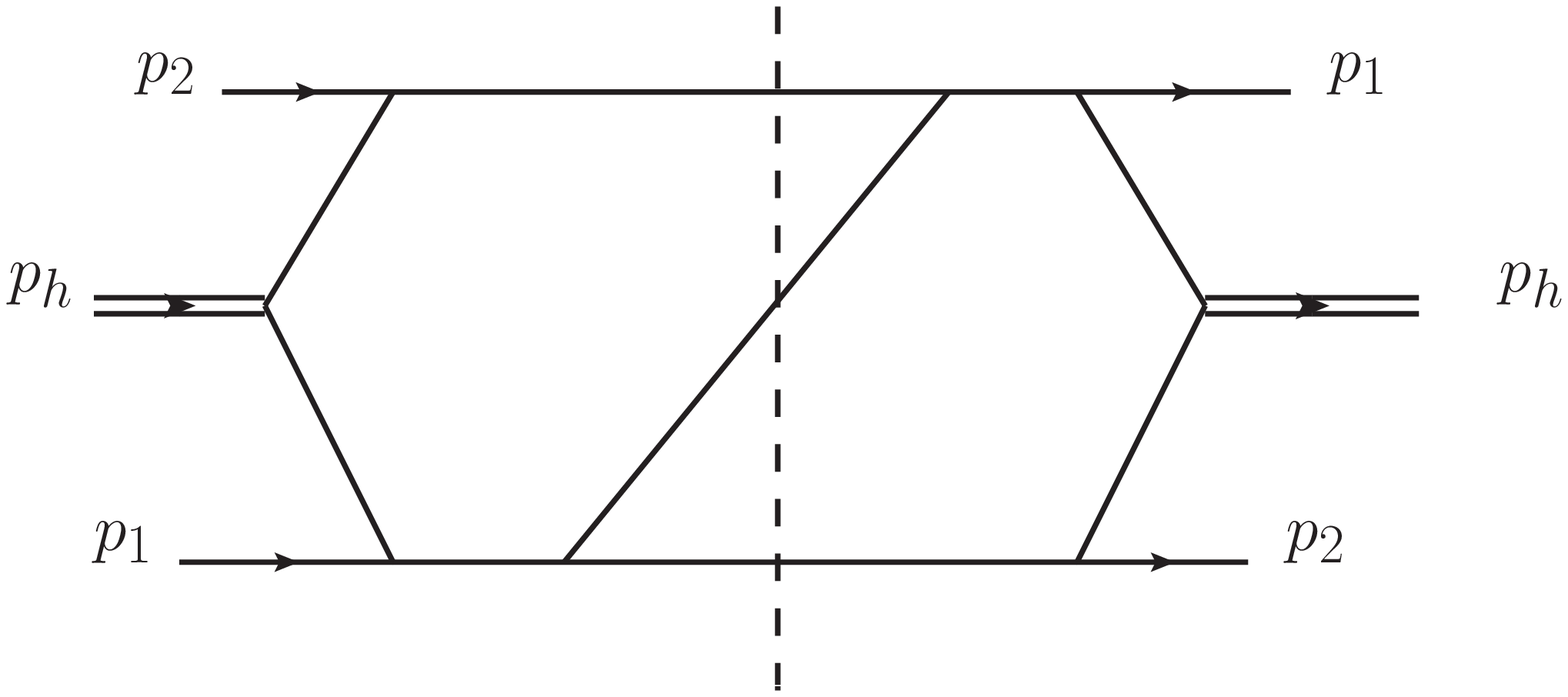}}\end{gathered} \nonumber\\[10pt]
&=&\int d\Phi^{0-3} \frac{1}{p_{23}^2p_{25}^2p_{34}^2p_{45}^2p_{134}^2p_{145}^2},\hspace{1cm} p_{i_1 \dots i_n}=p_{i_1}+\dots+p_{i_n}
\eea
In the above picture solid lines correspond to scalar propagators, doubled line to massive external legs and lines crossed by the dashed line represent the on-shell constraints of the phase space integration measure.\\
As described above, the threshold limit corresponds to the kinematic configuration where all radiation produced in association with the Higgs boson is uniformly soft. It is thus natural to perform the variable transformation $p_f\rightarrow \bar z p_f$~\cite{Anastasiou:2013srw}. Here, $p_f$ indicates the momentum of any final state parton, which can be identified in \eqref{RRRex} by the momenta crossing the dashed line, namely $p_{3}$, $p_4$ and $p_5$.
This rescaling induces a transformation on the phase space measure $d\Phi^{0-3} \rightarrow  z^{2d-6} d\Phi^{0-3}$ contained in \eqref{RRRex} given that $k\rightarrow \bar z k$.\\
Performing a series expansion of the integrand yields 
\bea
I(p_1,p_2,\bar z k)&=&\bar z^{2d-14} \left[\int d\Phi^{0-3} \frac{1}{(2p_2p_3)(2p_2p_5) p_{34}^2 p_{45}^2 (2p_1p_{34})(2p_1p_{45})} +\mathcal{O}(\bar z^1)\right]\nonumber\\
&=&\bar z^{2d-14} \left[ I^{(0)}+\bar z I^{(1)}+\dots\right].
\eea

Every term in the soft expansion of a Feynman integral can be written in terms of a linear combination of soft integrals as already observed in ref.\cite{Li:2014bfa,Anastasiou:2015yha}.
Soft integrals are Feynman integrals that are independently homogeneous under rescaling of the initial momenta $p_1$, $p_2$ or all momenta in the integral simultaneously.\\
Consider for example what happens to our example integral $I^{(0)}$ as we rescale $p_1\rightarrow \lambda_1 p_1$.
\beq
I^{(0)}\rightarrow \lambda_1^{\gamma_1} I^{(0)}.
\eeq
The associated rescaling dimension can be easily read off the integral and in the specific case of our example we find $\gamma_1=-2$.\\
In general,
	\bea
\label{eq:scaling}
p_1\rightarrow \lambda_1 p_1: &&\hspace{1cm} I_s\rightarrow \lambda_1^{\gamma_1} I_s.\nonumber\\
p_2\rightarrow \lambda_2 p_1: &&\hspace{1cm} I_s\rightarrow \lambda_2^{\gamma_2} I_s.\nonumber\\
\{p_1,p_2,p_f,k\} \rightarrow \lambda_3 \{p_1,p_2,p_f,k\} : &&\hspace{1cm} I_s\rightarrow \lambda_3^{\gamma_3} I_s.
\eea
The last line in the above equation indicates a simultaneous rescaling of all momenta in the curly bracket. 
Note that the respective scaling dimensions $\gamma_i$ depend on the specific integral in question but for simplicity we write them without any argument.

We realize that the integrated soft integrals are functions of four Lorentz invariant scalar products $s$, $k^2$, $2p_1k$ and $2p_2 k$. 
We can use the scaling behavior of our soft integrals to determine its functional dependence on three of the four scalar products. 
Consequently, the soft integrals depend on one variable that is invariant under any of the three rescaling symmetries. This invariant cross ratio is given by the dimensionless variable $x=\frac{k^2 s}{2kp_1 2kp_2}$, introduced in eq.~\eqref{eq:xsdiffhad2}.
Combining these properties we are able to write:
\beq
I_s(s,k^2,2kp_1,2kp_2)=s^{\gamma_1+\gamma_2-\gamma_3/2}(2kp_1)^{\gamma_3/2-\gamma_2}(2kp_2)^{\gamma_3/2-\gamma_1} \tilde{I}_s(x).
\eeq
Once our integrand is expressed in terms of integrals that only depend on the cross ratio $x$ we can use standard phase space integral techniques to compute these functions.
In particular, we employ the framework of reverse unitarity~\cite{Anastasiou2002,Anastasiou2003,Anastasiou:2002qz,Anastasiou:2003yy,Anastasiou2004a} to express our differential partonic cross sections in terms of a few soft master integrals.
Subsequently, we make use of the method of differential equations~\cite{Kotikov1991,Gehrmann2000,Henn2013} to compute our soft master integrals. 
The soft expansion greatly simplifies these steps as in the expression above we only need to maintain functional dependence on $x$. 
The resulting functions are harmonic polylogarithms~\cite{REMIDDI2000} in the $x$ variable.

Note, that so-far we did not apply the relation among the invariants that arises due to the on-shell constraint for the Higgs boson, $p_h^2-m_h^2=s+2kp_2+2kp_3+k^2-s z=0$. 
This relation is inhomogeneous under rescaling the final state momenta $k$ and will introduce sub-leading terms in the $\bar z$ expansion. 
In ref.~\cite{Anastasiou:2013mca} this issue was solved by performing a systematic expansion of the on-shell constraint by including it in the reverse unitarity method. 
Here, we choose to impose the aforementioned on-shell constraint only after reduction to master integrals.\\

An additional complication arises when loop integrals are part of the phase space integrand. 
The loop momentum can take arbitrary values that are parametrically smaller or larger compared to the parameter $\bar z$ we want to expand our cross section in. 
The obstacle is easily illustrated by regarding a single propagator containing a loop momentum $l$ and a rescaled final state momentum $\bar z p_3$.
The expansion 
\beq
\frac{1}{(l^2+\bar z 2 l p_3)}=\frac{1}{l^2}\sum\limits_{i=0}^\infty\left(-\frac{\bar z2 l p_3}{l^2}\right)^i,
\eeq
is simply not convergent if for example $l$ is uniformly smaller than $\bar z$.

To be able to perform a systematic threshold expansion at the integrand level for the partonic cross section we rely on the strategy of regions~\cite{Beneke:1997zp} that allows to consistently treat the problem. 
In particular, we want to expand contributions with one loop and two additional partons in the final state around the threshold limit. 
Exactly this issue was addressed in much detail in ref.~\cite{Anastasiou:2015yha} and we refrain from repeating the procedure here. 
Once the loop and phase space integrand is expanded we perform a reduction of the loop integrals to loop master integrals. \\
The initial step of performing the reduction of loop integrals allows us to determine the rescaling behavior of the loop integrals under the scaling transformations introduced in eq.~\eqref{eq:scaling}.
Next, we embed the loop master integrals again in terms of their Feynman propagator representation into the phase space integration and we subsequently perform the combined loop and phase space integral in the same fashion as the pure phase space integrals discussed above (i.e. via reverse unitarity and differential equations). 
Again, we benefit from having only to maintain functional dependence on the cross ration $x$ by inferring the dependence of our integral on the other variables from its behavior under scaling transformations.\\

With the techniques summarized in this section we can perform a threshold expansion of the partonic Higgs differential cross sections at arbitrary order in the strong coupling constant and to arbitrary power in $\bar z$.
Specifically, we perform the computation of tree level partonic cross sections with three partons in the final state and partonic cross section with one loop and two partons in the final state to first and second order in the threshold expansion. 
Extending the threshold expansion to higher powers is technically challenging and is left for future work. 
As a result we obtain all ingredients to compute the first and second term in the threshold expansion of the N$^3$LO Higgs differential cross section.

\section{Validating the Threshold Expansion for differential Observables at NNLO}
\label{sec:threshold}
In proton collisions with a fixed center of mass energy the probability for two constituent partons to collide can be understood as a function of center of mass energy of the colliding partons. 
This probability is falling as the center of mass energy of the colliding partons rises. 
In particular, if the two partons under consideration are gluons this probability is falling faster as a function of their energy than the probability for two quarks or one quark and one gluon. 
As the main source for the production of a Higgs boson at the LHC are two colliding gluons this implies that there is a kinematic enhancement for the Higgs boson to be produced from a system of gluons with as little energy as possible. 
The lowest possible energy to produce a Higgs boson is referred to as its production threshold and corresponds to the Higgs boson mass.  
In general, it can be expected that the bulk of Higgs bosons produced in proton collisions are produced at their production threshold and the cross section to find more energy in the produced system is falling with energy. 

In the past this simple kinematic consideration was exploited to derive simplifications for the prediction of Higgs boson cross sections. 
Perturbative NNLO~\cite{Harlander:2002wh} and N$^3$LO~\cite{Anastasiou:2015ema} corrections were approximated in the form of an expansion around the threshold. 
Factorization properties of the leading term in the threshold expansion are commonly exploited to perform all order resummation of threshold enhanced terms. 
In this section we will analyze the performance of an expansion of partonic Higgs differential cross sections around the production threshold.

As the partonic differential cross sections were computed analytically as a function of $\bar z$ in ref.~\cite{Dulat:2017aa} we can easily perform a threshold expansion a posteriori. In the following we want to study the quality of this approximation as higher and higher terms in the expansion are included for differential observables. 
In order to do so we compute the rapidity and transverse momentum distribution of the Higgs boson.
We perform a threshold expansion for all matrix elements occurring at NNLO and keep lower order matrix elements exact. 
We then truncate the expansion at different orders and compare with the exact results. 

To derive numerical values we numerically perform the remaining integrals in eq.~\eqref{eq:xsdiffhad2}  over the partonic cross sections in conjunction with MMHT2014 parton distribution functions~\cite{Harland-Lang:2014zoa} in a private C++ implementation. 
We perform renormalization and convolutions with the mass factorization counter terms numerically.
We only expand the partonic NNLO matrix elements around the production threshold. 
This leads to a mismatch in the cancellation of infrared and ultra violet divergences which are treated in the framework of dimensional regularization. 
Specifically the cancellation of poles in the dimensional regulator is only given up to the respective order in the threshold expansion at which we truncate. 
Throughout this section we choose the perturbative scale to be $\mu=m_h$.


\begin{figure*}[!ht]
\centering
\includegraphics[width=0.6\textwidth]{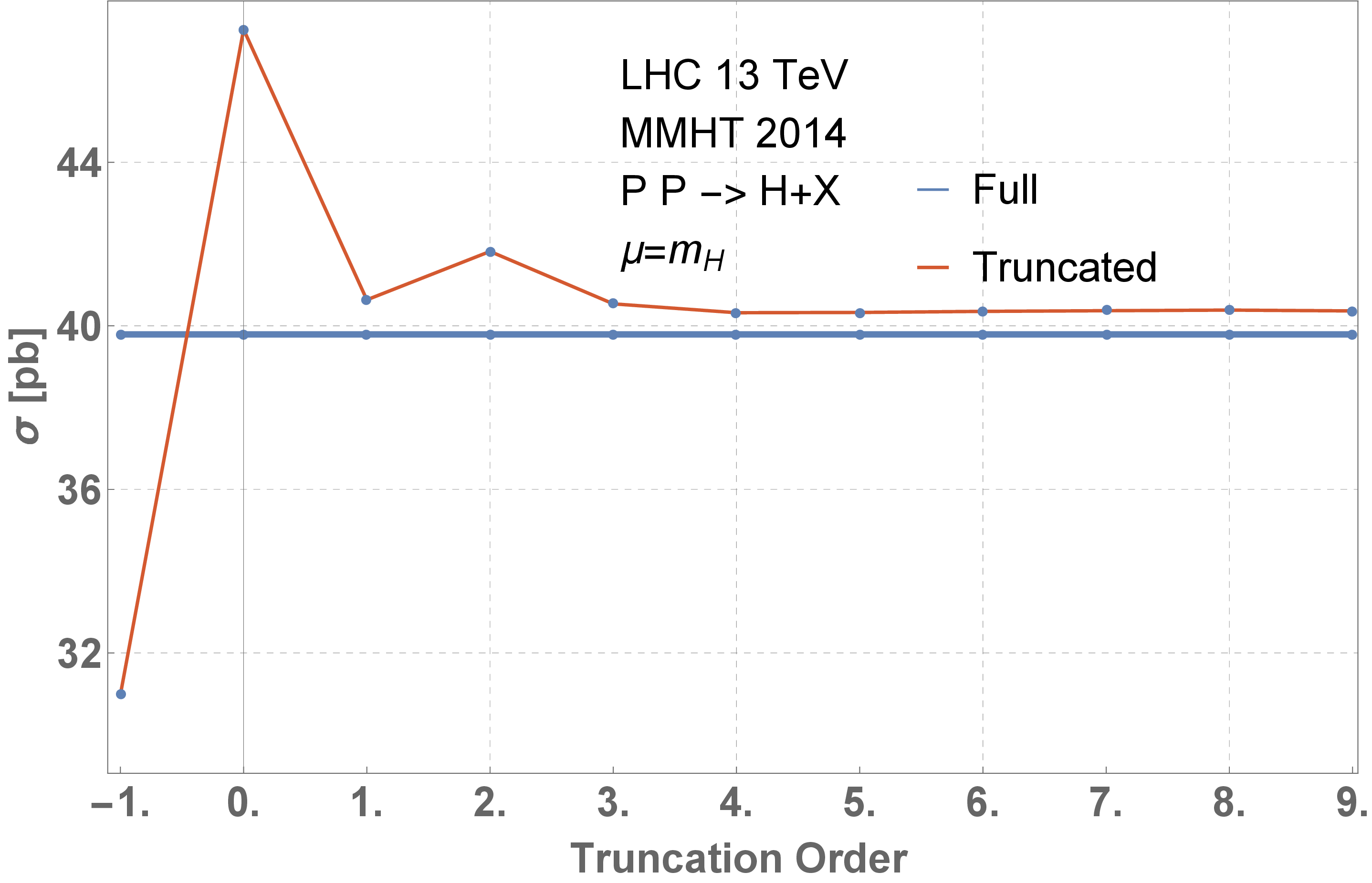}
\caption{Inclusive production probability for a Higgs boson as a function of the truncation order of the threshold expansion for the NNLO correction (red). The blue line represents the unexpanded NNLO inclusive cross section.
\label{fig:incxs}}
\end{figure*}

Let us first consider the inclusive cross section produced at a perturbative scale $\mu=m_h$. 
In refs.~\cite{Herzog:2014wja} similar studies were performed for the inclusive cross section at NNLO and our findings agree. 
We show the inclusive cross section through NNLO in fig.~\ref{fig:incxs} for different truncation orders. 
The first few terms in the threshold expansion (in red) significantly deviate from the exact result (in blue). After the first five terms the expansion stabilizes and subsequent terms gradually improve the result. 
The agreement after including five terms in the expansion is fairly good. Further improvement is achieved at a comparably slow rate. This slow convergence of the remaining difference to the exact result can be attributed to explicit divergences of the partonic coefficient functions at the high energy limit $z=0$. A similar behavior was observed for the expansion of the N$^3$LO corrections to the inclusive cross section in refs.~\cite{Anastasiou:2015ema,Anastasiou:2016cez,Herzog:2014wja}.


\begin{figure*}[!ht]
\centering
\begin{subfigure}[b]{0.45\textwidth}
\includegraphics[width=0.95\textwidth]{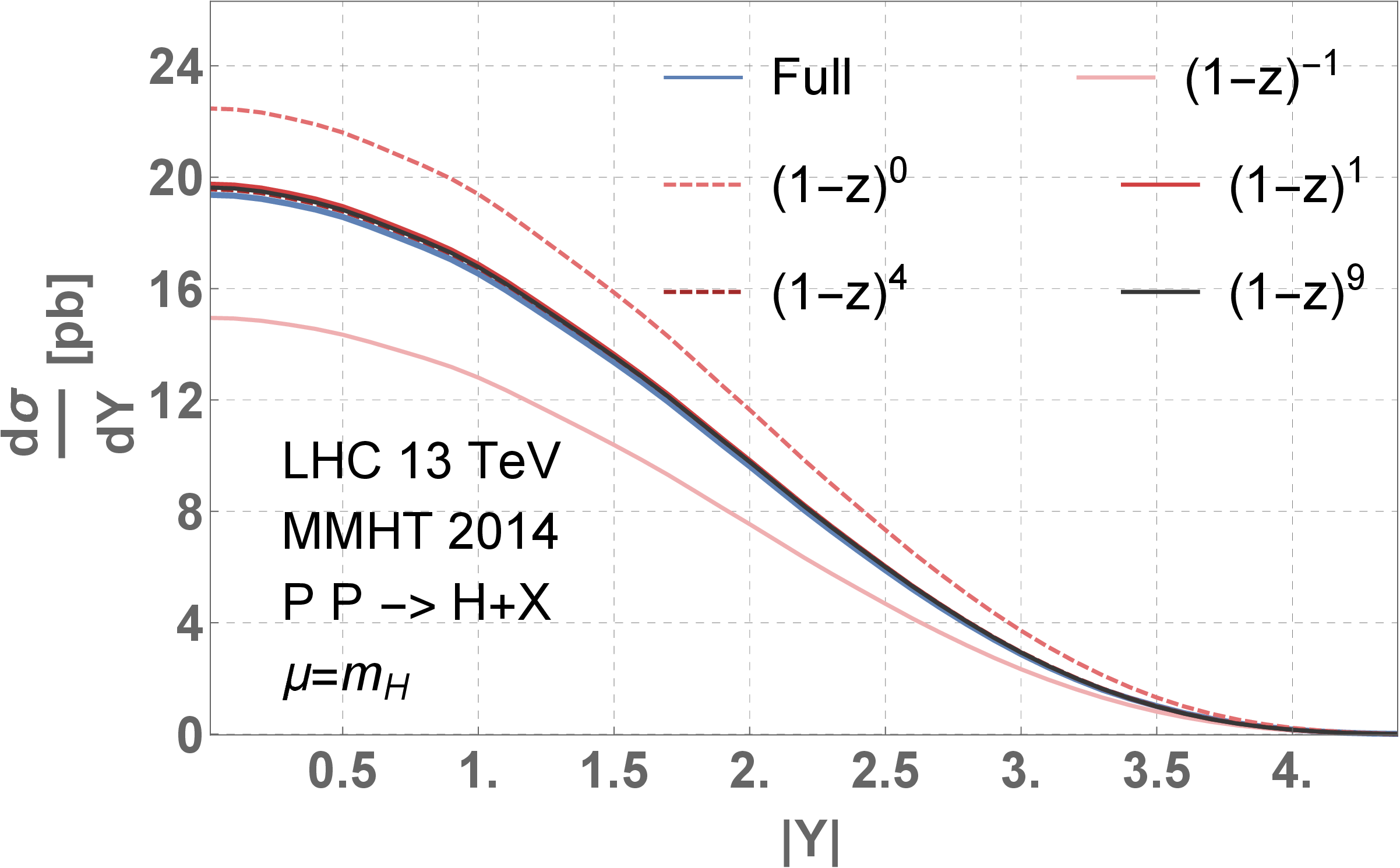}
\caption{
}

\label{fig:rapPhotInc}
\end{subfigure}
\begin{subfigure}[b]{0.45\textwidth}
\includegraphics[width=0.95\textwidth]{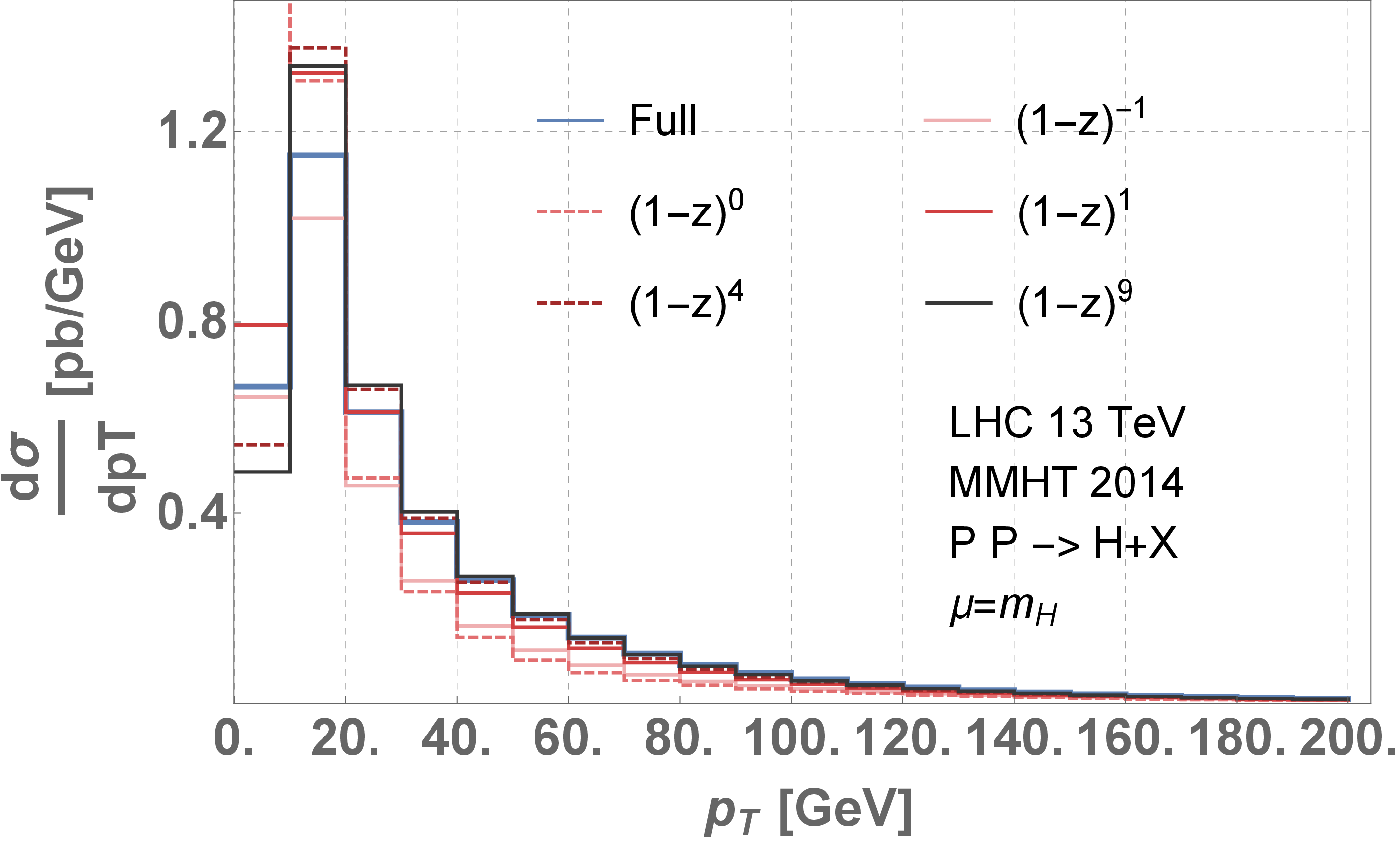}
\caption{
}
\label{fig:TH}
\end{subfigure}
\caption{\label{fig:THdiff1}NNLO absolute rapidity (left) and transverse momentum (right) distribution for the Higgs boson. The blue line represents the exact result. Increasingly darker shades of red represent higher and higher truncation order of the threshold expansion.}
\end{figure*}

In fig.~\ref{fig:THdiff1} we show the rapidity and transverse momentum distribution. 
Increasing truncation order of the threshold expansion is indicated by increasingly dark shades of red and the exact result in blue.
As for the inclusive cross section we observe that the first few terms display large variations from the full result. 
After about five terms the expansion stabilizes and adding higher terms shows gradual improvements on the approximation. 
Note that in order to derive a physical prediction for the first couple of bins in the transverse momentum distribution logarithms of the transverse momentum should be resumed to all orders in perturbation theory. No such procedure was applied here as this is beyond the scope of this article.


\begin{figure*}[!ht]
\centering
\begin{subfigure}[b]{0.45\textwidth}
\includegraphics[width=0.95\textwidth]{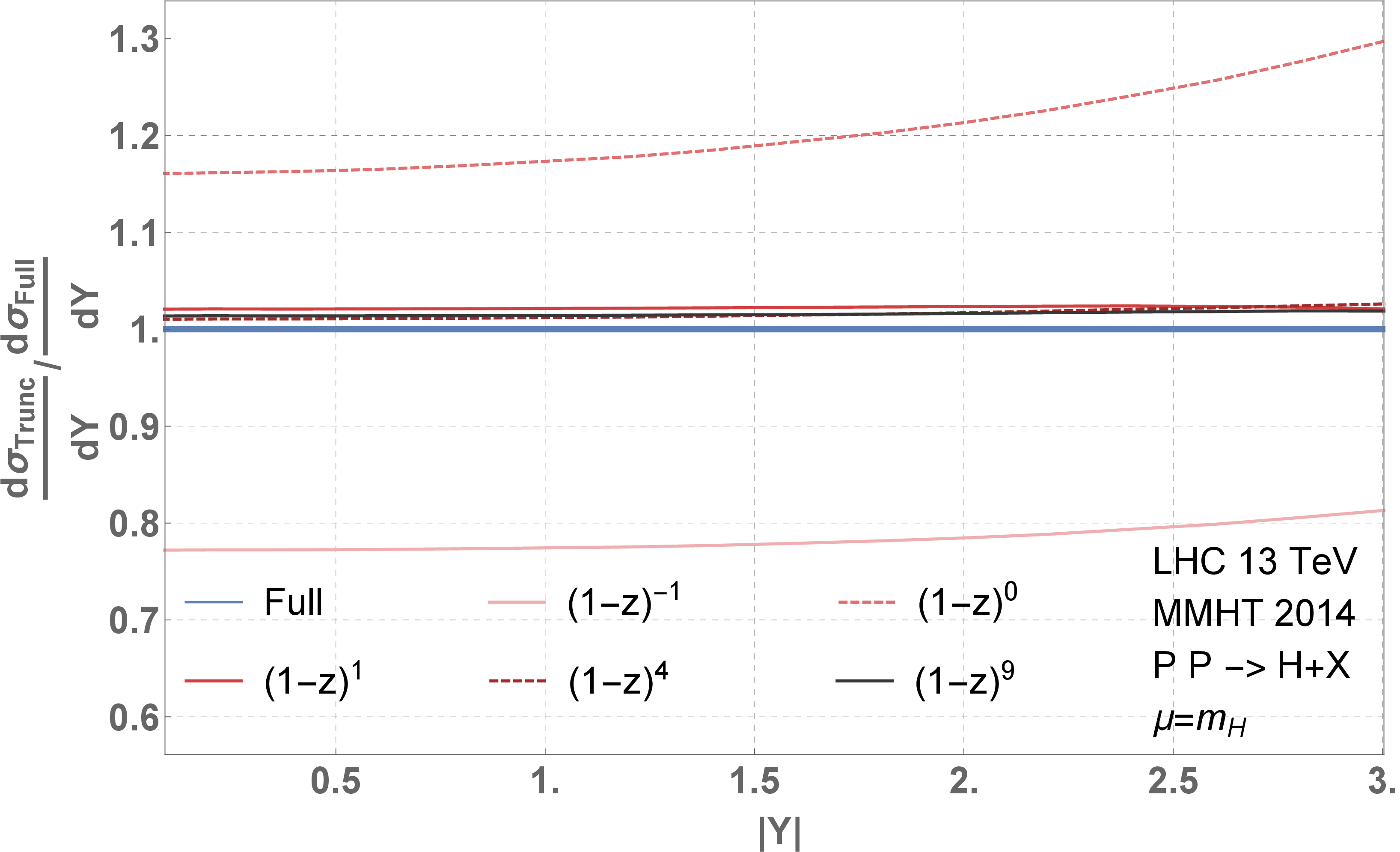}
\caption{
}
\label{fig:rapPhotInc}
\end{subfigure}
\begin{subfigure}[b]{0.45\textwidth}
\includegraphics[width=0.95\textwidth]{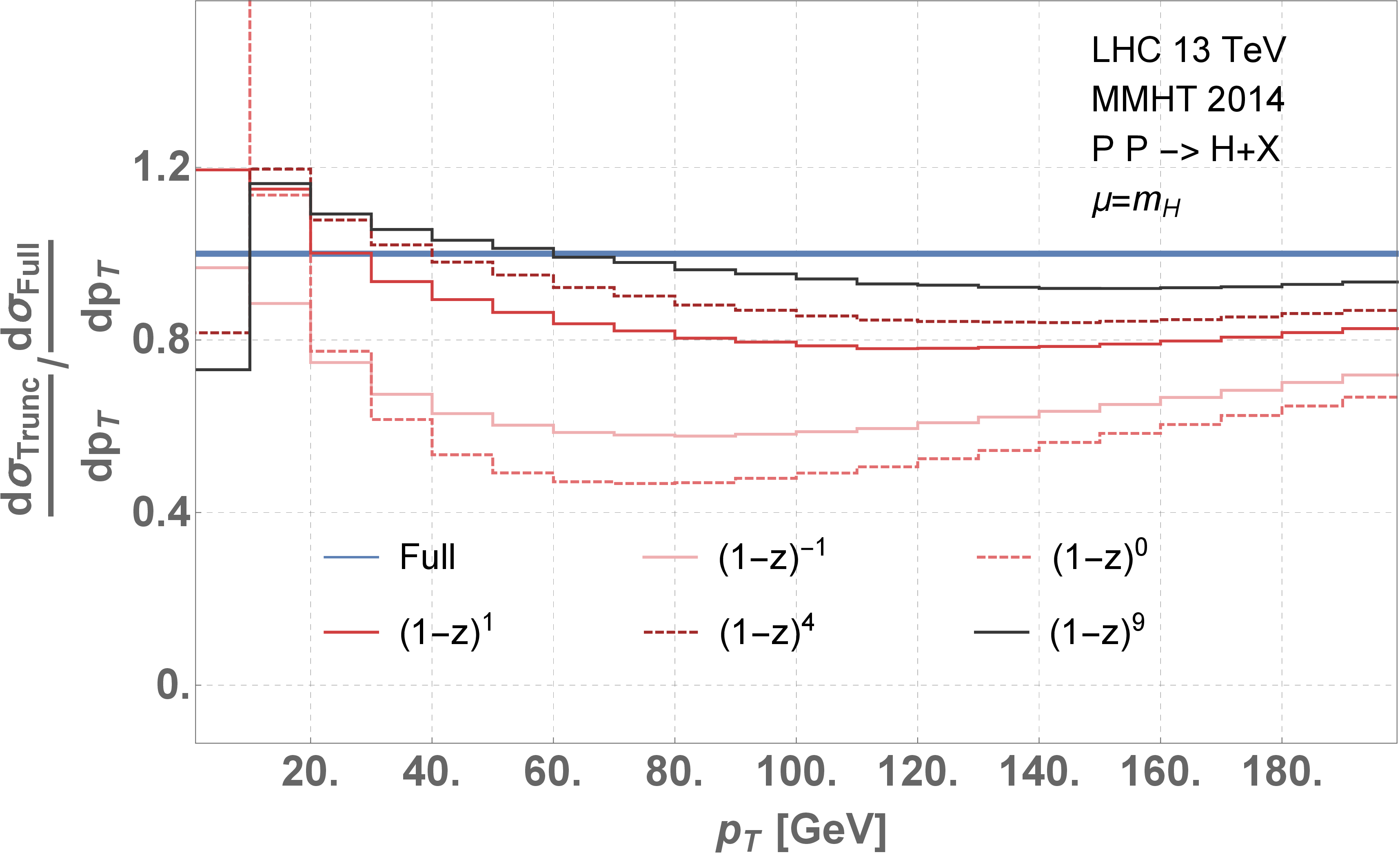}
\caption{
}
\label{fig:TH}
\end{subfigure}
\caption{\label{fig:THdiff2}NNLO absolute rapidity (left) and transverse momentum (right) distribution for the Higgs boson. The blue line represents the exact result. Increasingly darker shades of red represent higher and higher truncation order of the threshold expansion. All lines are normalized to the exact NNLO distributions.}
\end{figure*}

The actual quality of the expansion can be studied in more detail by analyzing the relative deviations of the expanded distributions from the full result. In fig.~\ref{fig:THdiff2} we show the rapidity and transverse momentum distribution normalized  to the unexpanded respective distributions. We note that by including only the third term in the expansion the rapidity distribution at NNLO is approximated to a level better than five percent. The transverse momentum distribution is improved as higher terms in the expansion are included. However, even with ten terms in the expansion the overall agreement between the exact result and the expansion is merely at the level of ten percent.

At large rapidity the quality of the threshold expansion deteriorates as the Higgs boson is produced with a larger boost along the beam axis and thus on average more energy is in the final state system.
The stark difference between the behavior of the rapidity and of the transverse momentum distribution can be understood by considering the structure of the partonic coefficient functions. 
The transverse momentum of the Higgs boson is identically zero at leading order as there is now parton produced for the Higgs boson to recoil against. At the kinematic threshold all radiation produced in association with the Higgs boson is soft and does not provide any recoil either. Adding terms in the threshold expansion only gradually builds up the functional dependence of the matrix elements on the transverse momentum. 
At the same time, the partonic matrix elements are singular as the transverse momentum of the Higgs boson vanishes. 
The partonic transverse momentum distribution contains kinematic singularities at finite values of $\bar{z}$ that are expanded by the threshold expansion. This leads to a slower convergence compared to the rapidity distribution.

We can conclude that the quality of a threshold expansion is subject to the particularities of individual observables. 
If such an expansion is to be used to approximate cross section predictions a dedicated analysis of the quality of the approximation has to be performed specific to every observable.
Complementing threshold expansions with expansions in the rapidity of the Higgs boson would be an interesting way forward and such studies are left for future work.\\


\begin{figure*}[!ht]
\centering
\begin{subfigure}[b]{0.45\textwidth}
\includegraphics[width=0.95\textwidth]{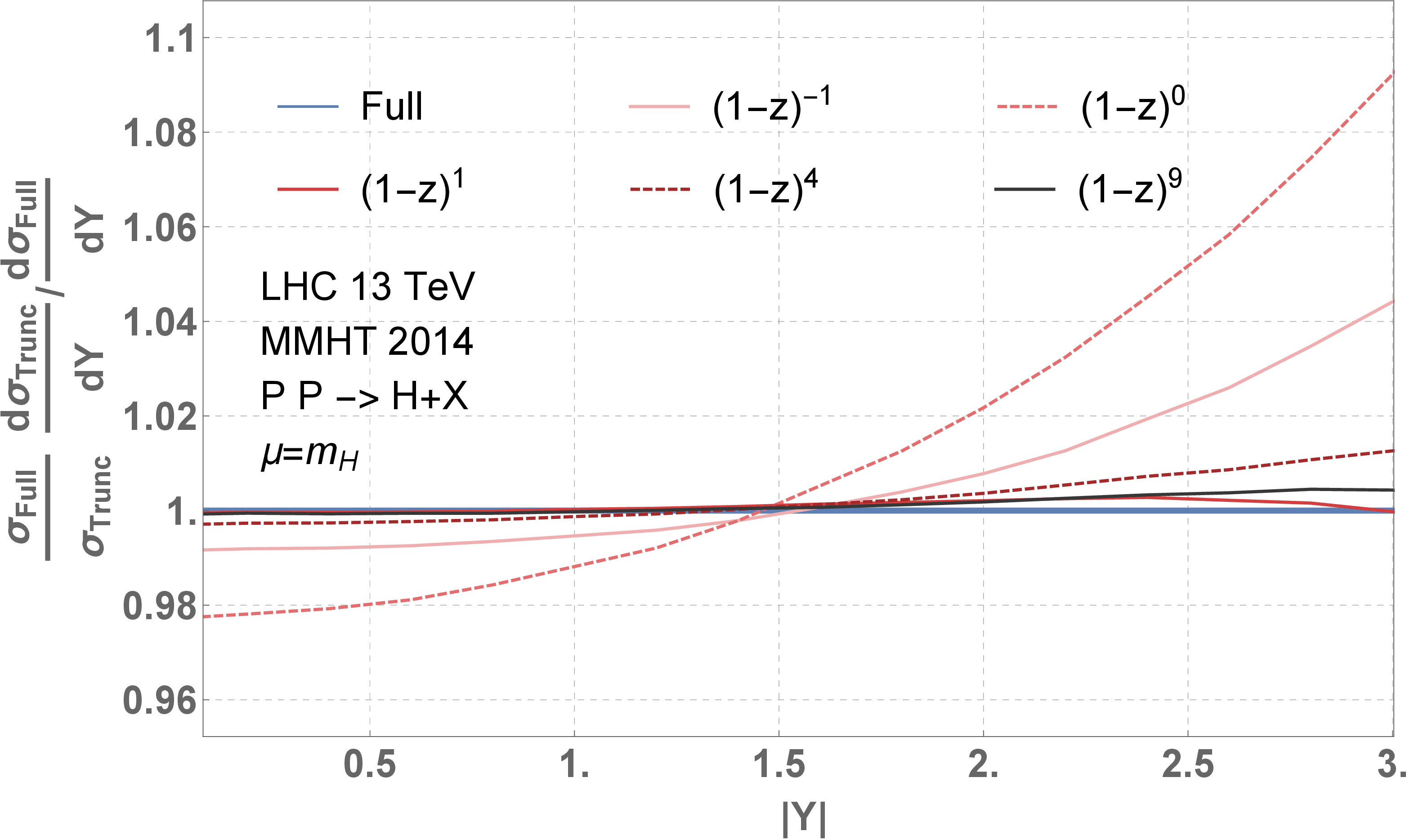}
\caption{
}
\label{fig:rapPhotInc}
\end{subfigure}
\begin{subfigure}[b]{0.45\textwidth}
\includegraphics[width=0.95\textwidth]{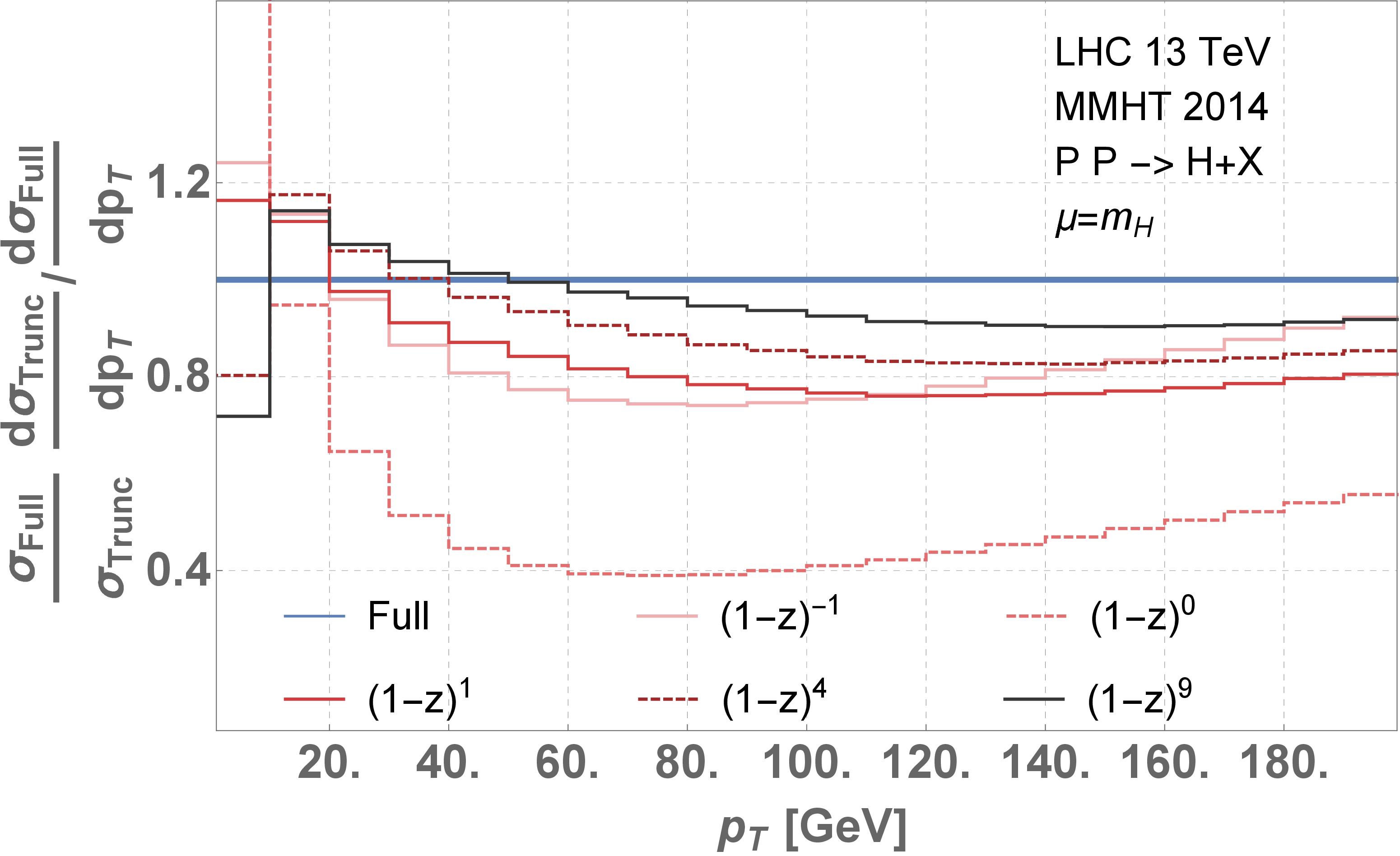}
\caption{
}
\label{fig:TH}
\end{subfigure}
\caption{\label{fig:THdiff3}NNLO absolute rapidity (left) and transverse momentum (right) distribution for the Higgs boson. The blue line represents the exact result. Increasingly darker shades of red represent higher and higher truncation order of the threshold expansion. All lines are normalized to the exact NNLO distributions. The expanded distributions were reweighted such that their cumulant reproduces the unexpanded NNLO cross section.}
\end{figure*}

It is commonly the case that predictions for inclusive cross sections become available prior to predictions for exclusive observables. 
Suppose this was the case for NNLO Higgs differential cross sections. In this case we could approximate differential cross sections at NNLO by performing a threshold expansion. To further improve the expanded results we could ensure that the inclusive cross section is reproduced by the differential approximation and only shapes are computed by the approximated result. We define the improved Higgs differential cross section to be 
\beq
\sigma^{\text{Improved}}_{PP\rightarrow H+X}\left[\O\right]=\frac{\sigma_{\text{Full}}}{\sigma_{\text{Expanded}}}\sigma^{\text{Expanded}}_{PP\rightarrow H+X}\left[\O\right].
\eeq
Here, $\sigma_{\text{Full}}$ and $\sigma_{\text{Expanded}}$ are the inclusive cross section without and with expanding around the threshold limit respectively. $\sigma^{\text{Expanded}}_{PP\rightarrow H+X}\left[\O\right]$ is the Higgs differential cross section based on partonic coefficient functions approximated by a threshold expansion. \\

We show predictions for the rapidity and transverse momentum distribution based on the improved approximation in fig.~\ref{fig:THdiff3} normalized to their exact counter parts. The effect of the rescaling is largest for low truncation order where the difference between the inclusive cross section based on the expansion and the full result is largest. We observe that the rapidity distribution is approximated at a level significantly better than two percent including just three terms in the threshold expansion throughout the rapidity interval $[0,3]$. 
The significant deviations of the shape of the transverse momentum distribution based on the threshold expansion are only mildly impacted by rescaling the distribution to the correct inclusive cross section.

\section{Numerical Results for approximate N$^3$LO Cross Sections}
In the following we discuss the remaining ingredients required to promote our analytical results for approximate N$^3$LO differential partonic cross sections to distributions. As in any calculation of hadronic observables we require parton distributions functions as an external input. Special care has to be taken not to introduce artifacts due to the interpolation of the input parton grids.
Next we discuss scale dependence of the cross section that we extract to all orders in the threshold expansion.
Combining all ingredients obtained, we then show differential distributions.

\subsection{Curious Encounters with Parton Distribution Functions}
\label{sec:PDFGames}
We compute predictions for the total cross sections and distributions by performing Monte-Carlo integrals over the remaining variables of the Higgs phase space given in eq.~\eqref{eq:xsdiffhad2}.
This requires numerical values for the parton distributions which are obtained by accessing the grids of standard PDFs~\cite{Dulat:2015mca,Harland-Lang:2014zoa,Ball:2014uwa,Alekhin:2017kpj,Butterworth:2015oua} through \texttt{LHAPDF}~\cite{Buckley:2014ana}.
The parton distributions functions $f(x,Q^2)$ are functions of the partonic momentum fraction $x$ and the scale $Q^2$.
Internally, the parton distributions are stored as finite grids in $(x,Q^2)$ space. These grids are interpolated on-the-fly by \texttt{LHAPDF} to provide the value of the parton distribution at the requested point in $(x,Q^2)$ space to the user. By default, this interpolation is performed using a log-cubic spline.

In the following, we want to study the numerical evaluation of the soft-virtual term of the rapidity distribution at N$^3$LO. In the strict soft limit, the rapidity distribution takes a particular simple form, since the partonic rapidity becomes unity. Thus the hadron level rapidity $Y$ is just a function of the parton momentum fractions, which can be seen from taking the limit $\bar{z}\to 0$ in eq.~\eqref{eq:svrap}.
The differential partonic cross section in that limit is therefore simply the soft-virtual inclusive cross section~\cite{Anastasiou:2014vaa} and we can write the hadronic cross section as,
\begin{equation}
  \begin{split}
    \label{eq:svrap}
      \frac{d\sigma_{gg}}{dY} = \frac{\sigma_0}{9v^2} \left(\frac{\alpha_s}{\pi}\right)^5 \int_{\tau}^{1}dz&f_g\left(\sqrt{\frac{\tau}{z Y}}\right)f_g\left(\sqrt{\frac{\tau Y}{z}}\right)\Bigg\{
      1124.31\,\delta(1-z)+1466.48\,\left[\frac{1}{1-z}\right]_+\\
      &-6062.09\left[\frac{\log(1-z)}{1-z}\right]_+
      +7116.02\,\left[\frac{\log^2(1-z)}{1-z}\right]_+\\
      &-1824.36\,\left[\frac{\log^3(1-z)}{1-z}\right]_+
      -230\,\left[\frac{\log^4(1-z)}{1-z}\right]_+
      +216\,\left[\frac{\log^5(1-z)}{1-z}\right]_+\Bigg\}\,.
  \end{split}
  \end{equation}

\begin{figure*}[!ht]
  \centering
  \includegraphics[width=1.00\textwidth]{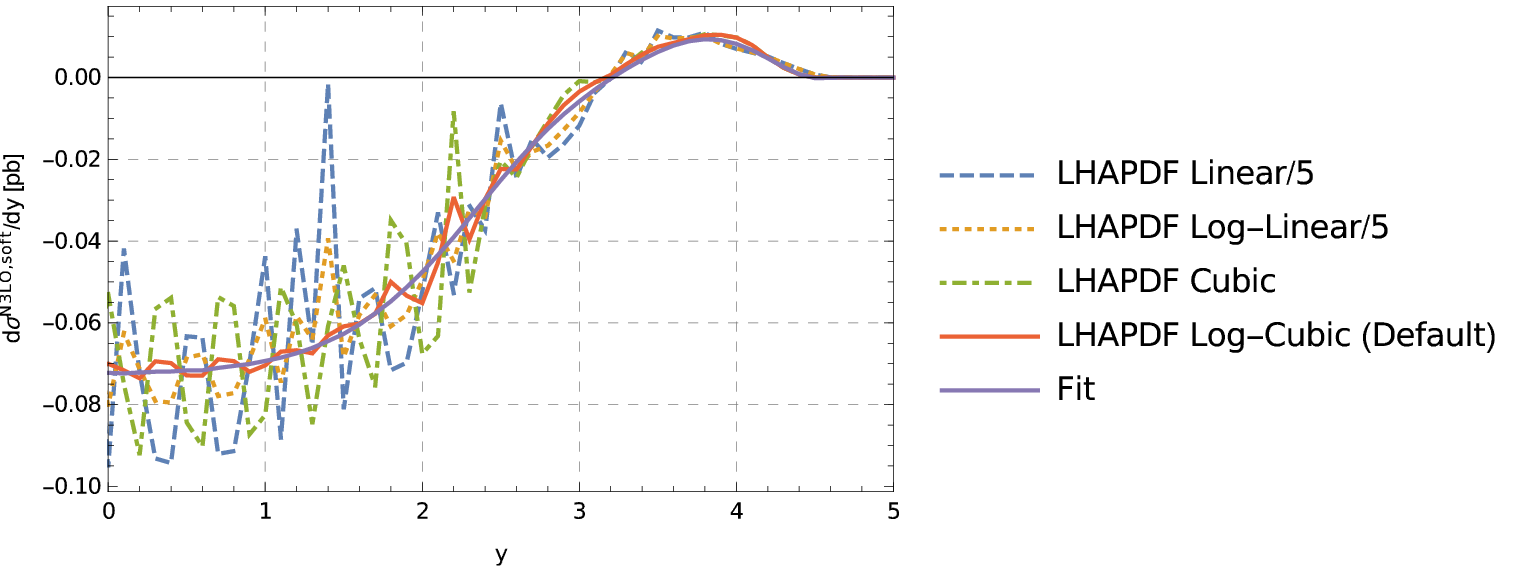}
  \caption{Soft-virtual term of the N3LO correction to the absolute rapidity distribution evaluated with various ways of interpolating the PDF grid.
    The blue line is obtained using a linear interpolator, the yellow line was obtained using a log-linear interpolator, the green line was obtained using a cubic interpolator. The solid orange line was obtained using the default \texttt{LHAPDF} setting, which uses a log-cubic interpolator. The solid purple line was obtained using our custom fit to the \texttt{LHAPDF} grid, which is described in the text. Note that the excursions of the blue and yellow line from the central value obtained through the fit were divided by a factor of five to be able to visualize the oscillations.}
  \label{fig:wiggles}
\end{figure*}

Evaluating the soft-virtual term of the rapidity distribution at N$^3$LO using the NNLO sets from \texttt{NNPDF} 3.0 at $\alpha_s(m_z)=0.118$ obtained using the default \texttt{LHAPDF} setup, we observe an unexpected loss of accuracy. As can be seen in fig.~\ref{fig:wiggles}, the rapidity distribution displays strong oscillations. Clearly, the soft-virtual limit of the rapidity distribution in eq.~\eqref{eq:svrap} has no structures that would warrant this oscillatory behaviour. Comparison with lower orders in perturbation theory suggest strongly that these features are numerical artifacts which should be suppressible by more careful numerics. The origin of these numerical artifacts can be understood when analyzing the influence of the interpolator used in LHAPDF to obtain continuous values from the discrete $(x,Q^2)$ grids.

Fig.~\ref{fig:wiggles} shows how the oscillations of the rapidity distribution change with different interpolation orders of the parton distributions. Clearly, these oscillations are artifacts of the interpolators lacking smooth higher order derivatives as can be seen from the fact that the magnitude of the oscillations increases when using lower rank splines for interpolation. These artifacts seem to be caused by the appearance of high powers of logarithms of $z$ in the partonic cross section. As can be seen in eq.~\eqref{eq:svrap}, the soft-virtual N$^3$LO contribution to the rapidity distribution contains terms of form
\begin{equation}
  \left[\frac{\log(1-z)^n}{1-z}\right]_+,
\end{equation}
for $n \in [0,5]$. In comparison, the soft-virtual NNLO contribution to the rapidity distribution contains only terms for $n \in [0,3]$. It seems therefore that the $n=4$ and $n=5$ terms pick up contributions from discontinuous higher moments of the interpolator.
This hypothesis can be tested at NNLO. With the default log-cubic interpolator the soft-virtual contributions to the NNLO rapidity distributions are smooth, as expected. Switching however to a log-linear interpolator, we see the same kind of oscillatory behavior appearing in the NNLO rapidity distribution.
It is clear that we need a smoother way to interpolate the \texttt{LHAPDF} grids in order to obtain useful predictions at N$^3$LO. One way to obtain smooth values from the parton distributions, is to evolve the parton distributions to a fixed value of $Q^2$ using \texttt{LHAPDF} and fit the resulting grid in $x$ space with an analytic function that is sufficiently smooth.
We make an empirical ansatz for the function as,
\begin{equation}
    \label{eq:pdffit}
  f_{Q^2}(x) = c_0(1-x)^{c_1}x^{c_2}+(1-x)^{c_3}\left[c_4+c_5\sqrt{x}+c_6 x +c_7 \log^2(x)+ c_8\log^4(x)+c_9\log^6(x)\right].
  \end{equation}
We fit this ansatz to points obtained from evolving the gluon NNLO \texttt{NNPDF} 3.0 grid to a scale $Q^2=(125\textrm{GeV})^2$, finding a $\chi^2/\textrm{ndof}$ of $1.9\times 10^{-7}$ with the parameters:

\begin{center}
  \begin{tabular}{r|r|r|r|r|r|r|r|r|r}
    $c_0$ & $c_1$ & $c_2$ & $c_3$ & $c_4$ & $c_5$ & $c_6$ & $c_7$ & $c_8$ & $c_9$ \\ \hline
    3.0752 & 4.7260 & 6.5836 & 3.7279 &-3.1264 &  7.9413 & -5.1894 & 0.4548 & -0.0004 & 0.0001
    \end{tabular}
  \end{center}

Evaluating the rapidity distribution with the fitted $x$-dependence for the PDFs, we obtain the smooth line in fig.~\ref{fig:wiggles}.

\begin{figure*}[!ht]
  \centering
  \includegraphics[width=1.00\textwidth]{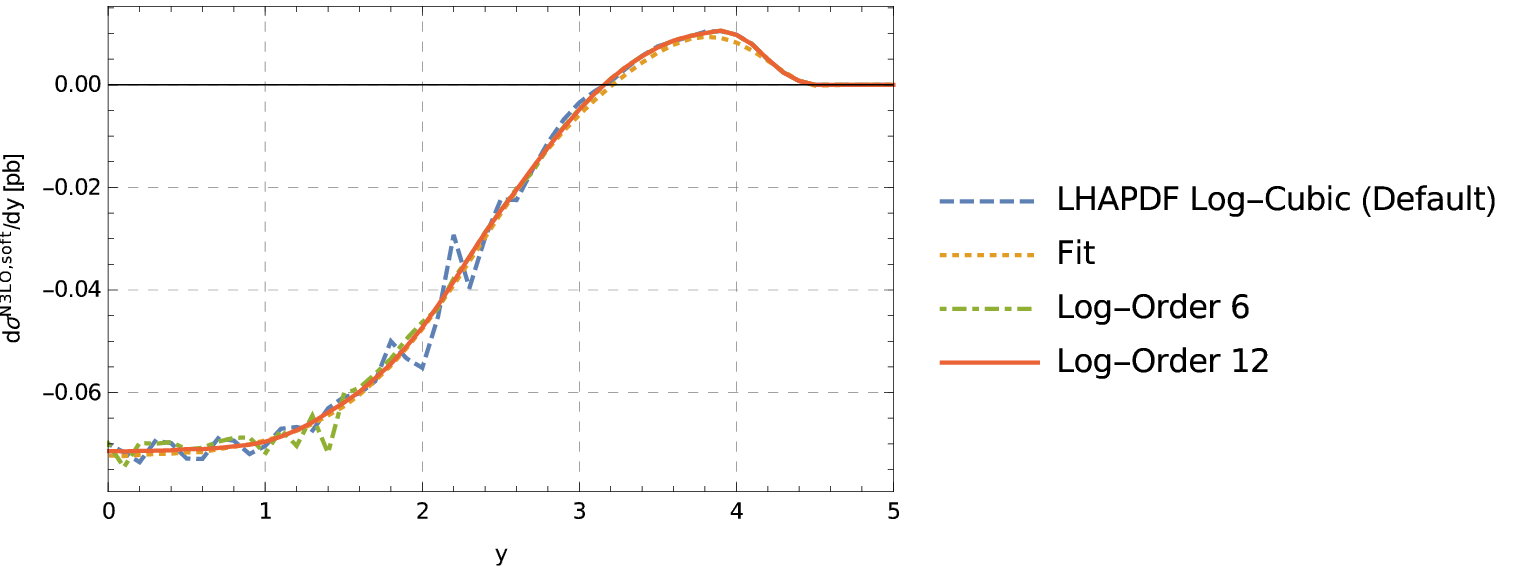}
  \caption{Soft-virtual term of the N3LO correction to the absolute rapidity distribution evaluated with various ways of interpolating the PDF grid. The dashed blue line shows again the default \texttt{LHAPDF} setup and the dashed yellow line shows again the fit for comparison. The dashed green line shows our interpolator using order 6 polynomials in $\log(x)$ space. The orange line shows the interpolator with order 12 polynomials in $\log(x)$ space.}
  \label{fig:smooth}
\end{figure*}

Performing a fit for parton distribution functions in terms of a smooth functions entails the disadvantage that this procedure has to be repeated for every required PDF set and ensuring a high fit quality sufficient for arbitrary observables is a non trivial task.
Furthermore, assessing the goodness of such a fit should be subsequently incorporated in the analysis of uncertainties of cross section predictions. 
As such it seems advantageous to use instead a higher order interpolator for  \texttt{LHAPDF} grids. 
We therefore implement a custom interpolator that is able to interpolate with splines of varying polynomial degree. 
We test the interpolation with degree six and degree twelve Legendre polynomials. The results can be seen in fig.~\ref{fig:smooth}.\\

As one can see in fig.~\ref{fig:smooth}, the interpolation with log-polynomial splines at order $6$ smoothes the oscillations in comparison to the default \texttt{LHAPDF} setup, however some artifacts still remain. Using log-polynomial splines at order $12$ then leads to acceptably smooth results.

The numerical results at N$^3$LO in the remainder of the paper are obtained using the order 12 interpolator. 

It should of course be noted that although the loss of accuracy cannot be directly observed in inclusive calculations at N$^3$LO, we can still expect an effect to appear from the integral over the oscillations.
To test this, we integrate the threshold expansion through 37 terms of the inclusive N$^3$LO cross section, as obtained in~\cite{Anastasiou:2015ema}, with two different ways of obtaining numeric values for the parton distributions.
We integrate the cross section with values for \texttt{NNPDF} 3.0 directly taken from LHAPDF and compare with integrating the cross section using the fit obtained in eq.~\eqref{eq:pdffit}.
We observe a deviation of about $1.3\%$ of the full N$^3$LO coefficient. We can therefore conclude that the effect on the total cross section through N$^3$LO is negligible.\\

We want to stress here the generality of the appearance of this loss of accuracy. Even though we show here numbers obtained using a particular PDF provider, we have investigated all common PDF sets from the big collaborations and find that these artifacts appear for any PDF set. We should also point out that the appearance of these oscillations is not specific to our calculation. These features had not been observed before, as the interpolators provided by default are smooth enough for NNLO calculations, however any calculation at N$^3$LO that relies on LHAPDF will be susceptible to this effect.
Clearly, it is desirable to study the effect of interpolator choices in more detail and to provide a flexible means of interpolating \texttt{LHAPDF} grids at sufficiently high orders to obtain smooth predictions for future N$^3$LO phenomenology.

\subsection{Exact Scale Variation at N$^3$LO}
\label{sec:scalevar}
After ultraviolet (UV) renormalization of the coupling constant and the Wilson coefficient, and after suitable redefinition of the parton distribution function, the Higgs differential cross section takes its final and finite form. 
The coefficient of $\alpha_S^3$ can be written as
\beq
\tilde \eta_{ij}^{(3)}(z,x,\lambda,L_\mu)=\sum\limits_{k=-3}^0\left[ \epsilon ^k \left(\frac{m_h^2}{\mu^2}\right)^{-3\epsilon} \eta_{ij}^{(3,k)}(z,x,\lambda) +\epsilon^k C_T^{(3,k)}\left(z,x,\lambda,L_\mu\right)\right]+\mathcal{O}(\epsilon).
\eeq
Here $L_\mu=\log\left(\frac{m_h^2}{\mu^2}\right)$.
The coefficients $C_T^{(3,k)}$ correspond to the Laurent series coefficients of the sum of UV renormalization counter term and mass factorization counter term.
They are constructed in the usual way in terms of lower order cross sections and universal anomalous dimensions and splitting functions~\cite{Dulat:2017aa}. 
The renormalized coefficient function is finite as the residual poles of the partonic coefficient function and the counter terms cancel. Consequently we find 
\beq
\eta_{ij}^{(3,k)}(z,x,\lambda)=-C_T^{(3,k)}(z,x,\lambda,0),\hspace{1cm}k<0.
\eeq
It is thus easy for us to construct these coefficients explicitly. Utilizing the above identity we may write the finite term of the N$^3$LO coefficient function as 
\bea
\tilde \eta_{ij}^{(3,0)}(z,x,\lambda,L_\mu)&=& \eta_{ij}^{(3,0)}(z,x,\lambda)+C_T^{(3,0)}\left(z,x,\lambda,L_\mu\right)+3C_T^{(3,-1)}\left(z,x,\lambda,0\right) L_\mu\nonumber\\
&-&\frac{9}{2}C_T^{(3,-2)}\left(z,x,\lambda,0\right) L_\mu^2+\frac{9}{2}C_T^{(3,-3)}\left(z,x,\lambda,0\right) L_\mu^3.
\eea
With this all contributions explicitly depending on the perturbative scale $\mu$ of the N$^3$LO coefficient functions are known. 
Additional dependence on the perturbative scale $\mu$ arises due to the multiplication of the partonic coefficient functions with the Wilson coefficient and due to the dependence of the strong coupling constant, the Wilson coefficient and the parton distribution functions on the perturbative scale. 

We now present the impact of the all contributions of the $\alpha_S^3$ coefficient on the Higgs differential cross sections. 
Specifically, we compute contributions to the rapidity distribution for the Higgs boson given by all ingredients that explicitly contain a logarithm $L_\mu$.
We include the partonic coefficient function at N$^3$LO as 
\beq
\tilde \eta_{ij,\,\,\text{RGE}}^{(3,0)}(z,x,\lambda,L_\mu)=\tilde \eta_{ij}^{(3,0)}(z,x,\lambda,L_\mu)-\tilde \eta_{ij}^{(3,0)}(z,x,\lambda,0)
\eeq
as well as all contributions to the $\alpha_S^3$ coefficient of the cross section containing renormalization group logarithms that arise due to the multiplication of the Wilson coefficient with lower order coefficient functions. 
We thus obtain all contributions to the N$^3$LO correction to the Higgs differential cross section involving explicit RGE logarithms.\\
\begin{figure*}[!ht]
\centering
\includegraphics[width=0.75\textwidth]{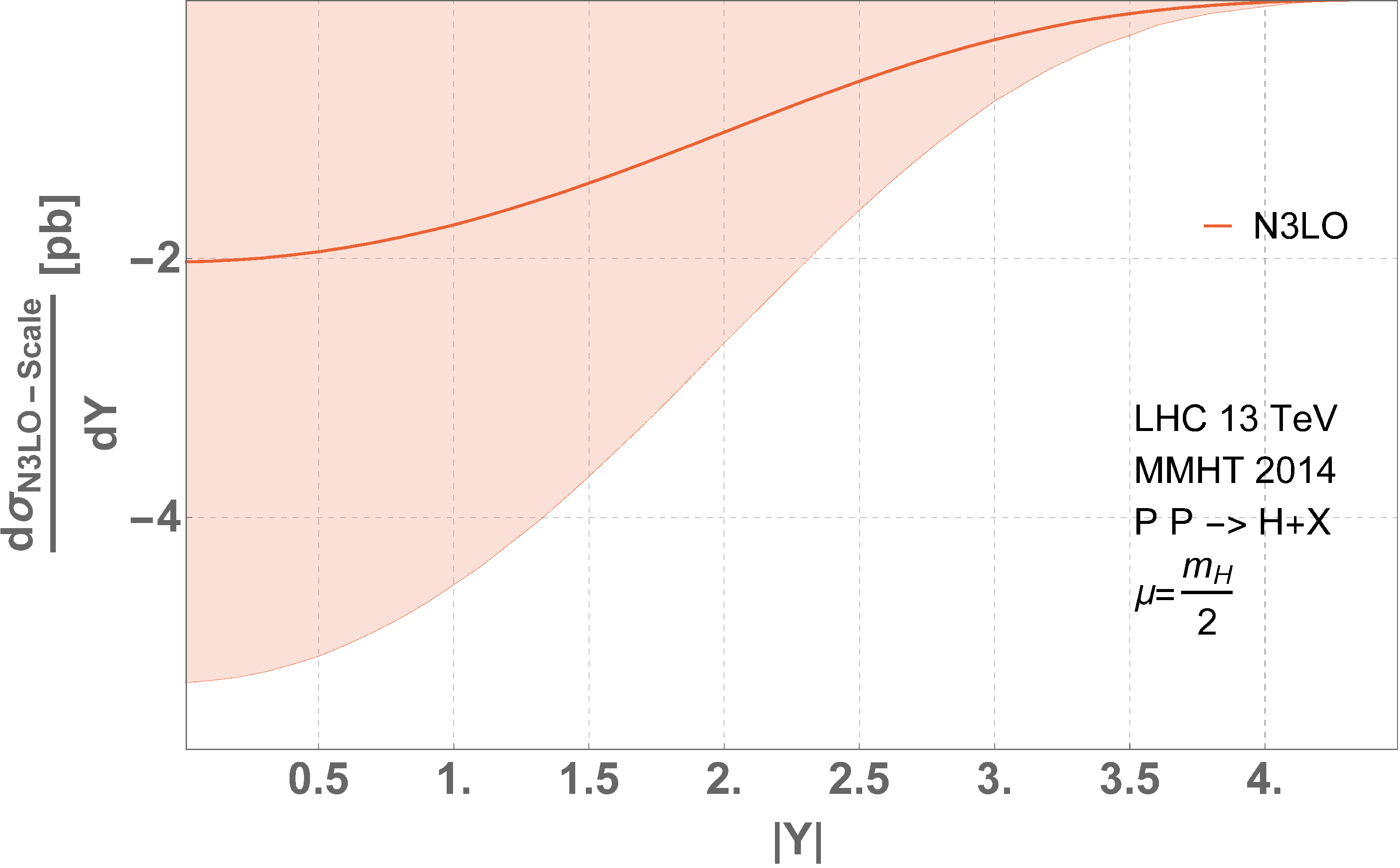}
\caption{\label{fig:scalevar}N3LO correction to the absolute rapidity distribution arising from coefficients of logarithms of the perturbative scale $\mu$. 
The band corresponds to a variation of the scale within the interval $\mu\in[m_h/4,m_h]$.}
\end{figure*}

We show numerical impact of the contributions involving RGE logarithms on the N$^3$LO corrections to the rapidity distribution of the Higgs boson in fig.~\ref{fig:scalevar}.
We choose $\mu=\frac{m_h}{2}$  as a central scale which results in the prediction given by the solid line. The red bands correspond to a variation of the perturbative scale in the interval $\mu\in\left[\frac{m_h}{4},m_h\right]$.
The contribution is monotonously rising as we increase the perturbative scale. 
At $\mu=m_h$ the argument of the RGE logarithm is one and the contribution considered here is identically zero.
The corresponding inclusive cross section agrees for all scales with the results of ref.~\cite{Anastasiou:2015ema}.
The contribution presented here on its own does not allow for an improved prediction at N$^3$LO since the finite coefficient functions without any RGE logarithms are still missing.
However, it represents one more essential stepping stone towards Higgs differential cross sections at N$^3$LO.

\subsection{Numerical Results for approximate differential Distributions at N$^3$LO}
\label{sec:N3LONumerics}
In section~\ref{sec:threshold} we discussed the phenomenological implications of performing a threshold expansion for Higgs differential cross sections at NNLO. 
The findings clearly indicate that several terms in the expansion are required. 
Particularly, predictions made by performing the threshold expansion to only the first or second order displayed sizable deviations from the true result.
Nevertheless, in section~\ref{sec:N3LOThresholdCalc} we went on to compute the first and second term in the threshold expansion of the N$^3$LO coefficient function. 
The main motivation is that this result provides key ingredients for the full analytic computation of the N$^3$LO coefficient functions. 
Furthermore it represents the complete soft counter term for the Higgs differential cross section at N$^3$LO. 
In this section we will demonstrate that the same pattern as observed for the first two terms in the expansion at NNLO proliferates at N$^3$LO.

\begin{figure*}[!ht]
\centering
\begin{subfigure}[b]{0.45\textwidth}
\includegraphics[width=0.95\textwidth]{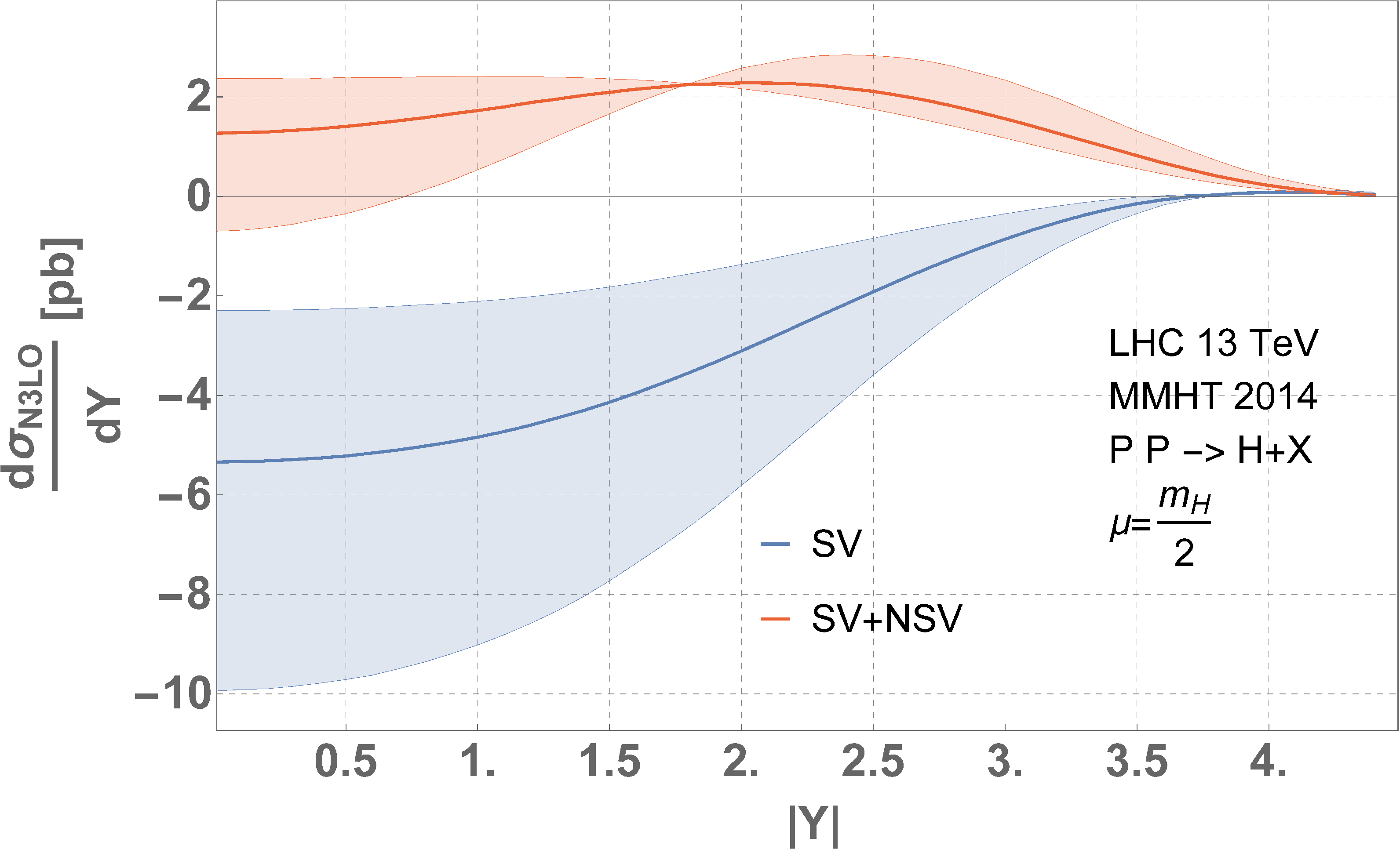}
\caption{
\label{fig:RapOnly}
}
\end{subfigure}
\begin{subfigure}[b]{0.45\textwidth}
\includegraphics[width=0.95\textwidth]{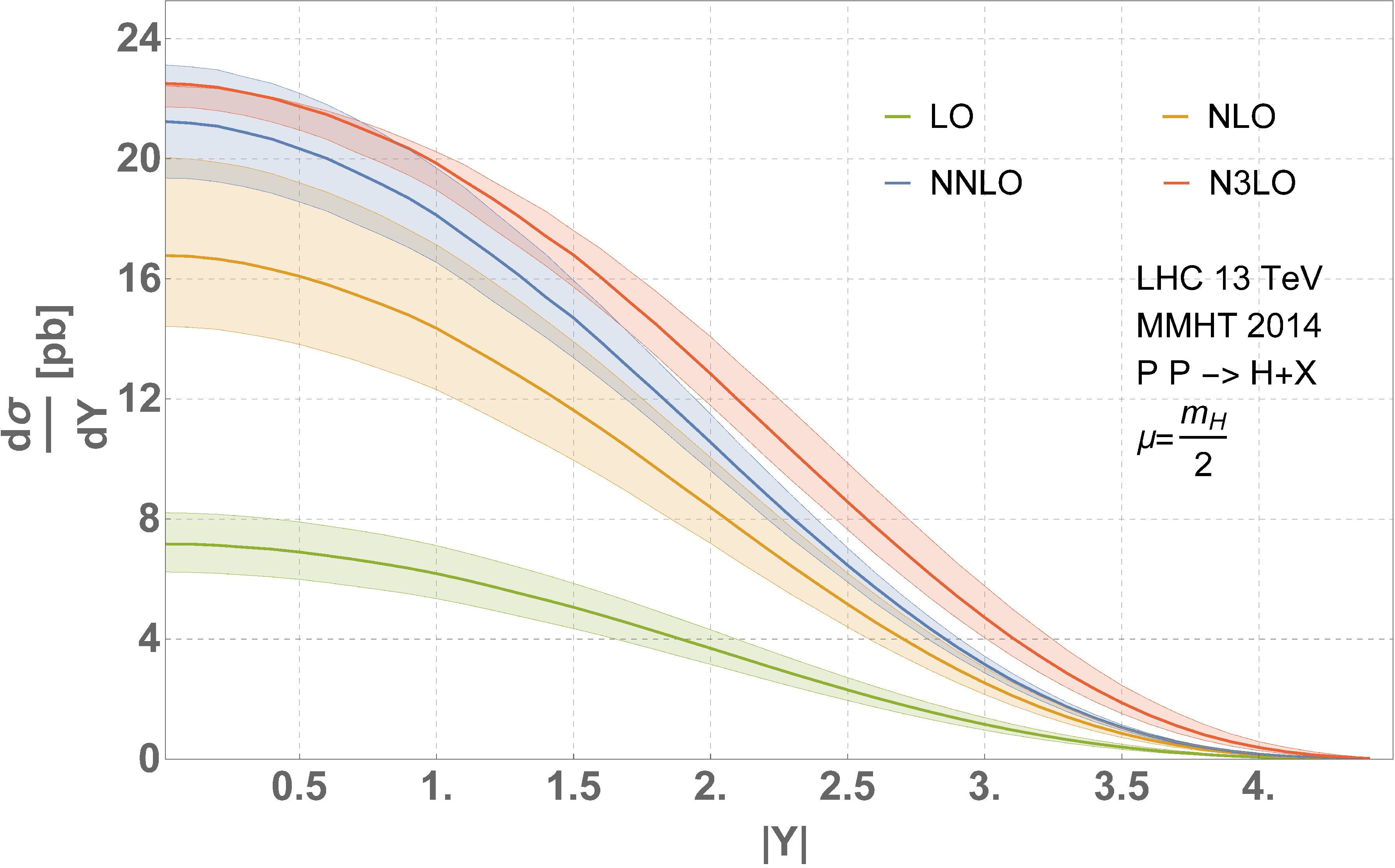}
\caption{
\label{fig:Rap}
}
\label{fig:TH}
\end{subfigure}
\caption{The left plot shows the contribution of the N$^3$LO partonic cross section to the absolute rapidity distribution of the Higgs boson approximated by including the first (blue) an by including the first and second (red) term in the threshold expansion. 
The right plot shows the rapidity distribution of the Higgs boson computed through different orders in perturbation theory. 
N$^3$LO contributions were approximated by performing a threshold expansion through the second term. }
\end{figure*}

We implemented the analytical results we obtained for the partonic coefficient functions in terms of the first and second term in the threshold expansion into a private 
c++ code. 
Furthermore, we combine our new results with our computation of the exact scale variation contributions obtained in section~\ref{sec:scalevar}. 
We show our results for the N$^3$LO corrections to the rapidity distribution of the Higgs boson in figure~\ref{fig:RapOnly}. 
Results including only the first term in the threshold expansion are shown in blue and including also the second term in red.
The bands correspond to variations of the perturbative scale around the central value $\mu=\frac{m_h}{2}$ by a factor of two. 
It is evident that the two predictions based on the first and second order expansion wildly differ which confirms our expectation from the analysis at NNLO. 
While the scale variation of the correction to the rapidity distribution based on the leading term in the threshold expansion is monotonously increasing with the scale the second order approximation is not.
The negative contributions arising from the explicit RGE logarithms are, depending on the exact value of the scale, compensated by the positive contributions arising from the N$^3$LO coefficient function and the Wilson coefficient.
Clearly, the scale variation can in no way describe the uncertainty due to truncation of the
    threshold expansion after a finite number of terms. If we were to derive phenomenological
    predictions from the threshold expansion at N$^3$LO, especially with only so few terms, we would
    have to carefull study the progression of the threshold expansion and derive a measure of
    uncertainty from e.g. analyzing the threshold expansion at an order where the full result is
    known, similar to what was done in~\cite{Anastasiou:2016cez} for the inclusive cross section.

In figure~\ref{fig:Rap} we combine the predictions for the corrections to the rapidity distribution at N$^3$LO based on the first and second term in the threshold expansion with lower order results (in red).
Exact lower order results are shown for LO, NLO and NNLO in green, yellow and blue respectively. 
We observe a fairly large impact of the approximate N$^3$LO corrections on the rapidity distributions.

The  inclusive cross section obtained with our current next-to-soft coefficient functions differs significantly from the inclusive cross section obtained in ref.~\cite{Anastasiou:2015ema}.
Our approximate results show large differences between the two newly computed terms. 
This confirms our NNLO analysis that demonstrates that an approximation based only on the first and second term in the threshold expansion is insufficient in order to improve on currently existing phenomenological predictions.
Further subleading terms in the threshold expansion or a complete computation are required. Improvements towards this goal will be part of future work.

\section{Conclusions}
\label{sec:conclusions}
In this article we achieve several key steps towards predicting differential observables to N$^3$LO in QCD perturbation theory.
We illustrate how a systematic expansion around the production threshold of Higgs-differential cross sections can be performed to arbitrary order. 
We apply this method to obtain the first and second term in the threshold expansion of the N$^3$LO coefficient functions in analytical form. 
This analytic data represents a corner stone of a complete N$^3$LO calculation as it constitutes the complete soft limit of the cross section and contains vital boundary information for the computation of master integrals via differential equations.
Furthermore, the obtained information may in future work serve as data to extract anomalous dimension for the resummation of logarithms in the transverse momentum of the Higgs boson.

Furthermore, we analyze the performance of a threshold expansion for the Higgs-differential cross section.
We start by considering corrections at NNLO, for which also the exact result is known. 
Analyzing the analytic structure of the NNLO coefficient functions shows that their threshold expansion is convergent within the unit interval of the threshold parameter $\bar z$. 
When studying the inclusive cross section in the threshold expansion, we observe that a oscillatory behaviour of the series when only including the first three powers in $\bar{z}$.
The series then stabilizes within inclusion of the first five coefficients, resulting in a difference of $3\%$ compared to the full NNLO result. 
Further improvement due to including even higher order terms is comparably slow.

Next, we analyze the quality of differential predictions obtained with threshold expansions.
The rapidity distribution of the Higgs boson at NNLO displays similar behaviour as the inclusive cross section.
Including about five terms stabilizes initial oscillatory pattern and leads to good approximations of the full result.
The distribution starts to deviate from the full result at high rapidities as more and more energy is required in the final state.
The second observable we analyze is the transverse momentum distribution of the Higgs boson.
The quality of the approximation obtained including the same amount of terms in the threshold expansion as for the rapidity distribution is greatly reduced.
While including higher and higher terms in the expansion is improving the approximation the convergence is so slow that even with ten terms in the expansion the deviations from the exact result are the level of ten percent.
In general, the approximations based on threshold expansions can be improved by normalizing differential cross sections such that their cumulant reproduce the exact inclusive result.\\
Our analysis at NNLO shows that the threshold expansion for Higgs-differential cross sections can be a powerful tool. 
The quality of the approximation has to be carefully assessed for every observable under consideration. 
Even for comparatively inclusive observables as the total cross section or the rapidity distribution several terms in the threshold expansion are required to obtain a reliable approximation.

We study the numerical impact of the newly obtained terms in the threshold expansion of the N$^3$LO coefficient function. 
The resulting rapidity distribution displays a similar pattern as we observed for the corresponding NNLO coefficient function. 
For improved phenomenological predictions more terms in the threshold expansion are required. 
Already now we obtain the full corrections at N$^3$LO due to terms with explicit dependence on the perturbative scale.

When computing corrections at N$^3$LO to the rapidity distribution of the Higgs boson we observe that the widely used framework for parton distribution functions \texttt{LHAPDF} needs to be modified.
The routines used by the tool to interpolate underlying grids for the parton distributions are insufficient to produce smooth distributions at N$^3$LO. 
As a consequence we observe an oscillatory pattern that is modulating the N$^3$LO correction to the rapidity distribution obtained with the soft-virtual approximation.
We advocate to implement a log-polynomial interpolator of order twelve or a smooth fitting procedure.

\section*{Acknowledgments}
We are grateful to Stefan Hoeche for illuminating discussions. 
We also thank Babis Anastasiou for fruitful discussions and useful comments on the manuscript.
AP is supported by the ETH Grant ETH-21 14-1 and the Swiss National Science Foundation (SNSF) under contracts 165772 and 160814. 
The work of FD is supported by the U.S. Department of Energy (DOE) under contract DE-AC02-76SF00515. 
BM is supported by the European Commission through the ERC grant  HICCUP.

\appendix
\section{Regularization of Coefficient Functions}
\label{sec:APPReg}

Consider a function $f(x) = x^{-1+a\epsilon}f_h(x)$, for some integer $a$ and with $f_h(x)$ holomorphic around $x=0$.
We are interested in integrating the function over a test function $\phi(x)$ on the range $[0,1]$.
In the case of our Higgs-differential cross section, the test function $\phi(x)$ corresponds to the product of the parton luminosity and the measurement function.
We can explicitly subtract the divergence at $x=0$ and integrate by parts to obtain
\bea
I&=&\int_0^1 dx f(x) \phi(x)= \int_0^1 x^{-1+a\epsilon} f_h(x) \phi(x)\nonumber\\
&=&\int_0^1 dx x^{-1+a\epsilon} \left[f_h(x)\phi(x)-f_h(0) \phi(0)\right] +\frac{1}{a\epsilon} f(0) \phi(0).
\eea
We now want to give an expression for the partonic cross section that is finite even if all inclusive integrations are performed.
To this end we define in a slight abuse of notation,
\beq
f_s(0)\equiv\delta(x)\left[x^{-1+a\epsilon}-\frac{1}{a \epsilon}\right]f_h(0).
\eeq
Here the $\delta$ distribution is to be understood as acting only on the test function and not on its coefficient in the square bracket.
It is easy to see that $f_s(0)$ integrates to zero.
We can therefore regulate the integrand $f(x)$ by subtracting $f_s(0)$,
\beq
I=\int_0^1 dx f(x) \phi(x)=\int_0^1 dx\,(f(x)- f_s(0))\,\phi(x),
\eeq
so that every term of its $\epsilon$ expansion can be integrated numerically.

In the case of our Higgs-differential cross sections, we need to regulate potential end-point divergences in the three remaining variables $\zb, x$ and $\lambda$,
c.f.~eq.~\eqref{eq:xsdiffhad2}.
We define the distributions $\sigma_s$ that subtract the limits of $\sigma(\zb, x, \lambda)$
and label them by the kinematic limit of the cross section that they reproduce.
For example $\sigma(\zb,0,\lambda)$ takes care of the limit of the cross section as $x$ goes to zero.
After partial-fractioning to avoid simultaneous singularities on both endpoints of the integral, we obtain the following decomposition, 
\begin{align}
\label{eq:finitexs}
\sigma_f(\zb ,x,\lambda)\equiv & \,\sigma(\zb ,x,\lambda)-\sigma_s (\zb,x,1)-\sigma_s (\zb,x,0 )-\sigma_s (\zb,1,\lambda)-\sigma_s (\zb,0,\lambda)-\sigma_s (0,x,\lambda)\nonumber\\
+&\,\sigma_s (\zb,1,1)+\sigma_s (\zb,1,0)+\sigma_s (\zb,0,1 )+\sigma_s (\zb,0,0)+\sigma_s (0,x,1)+\sigma_s (0,x,0)\nonumber\\
+&\,\sigma_s (0,1,\lambda)+\sigma_s (0,0,\lambda )-\sigma_s (0,1,1)-\sigma_s (0,1,0)-\sigma_s (0,0,1)-\sigma_s (0,0,0).
\end{align}

One main result of this article is the analytic computation of the partonic coefficient functions $\eta^{(k)}_{ij}(z,x,\lambda)$ as defined in eq.~\eqref{eq:etas}.
We created finite versions of this coefficient functions in the spirit discussed above
and provide them in \texttt{Mathematica} readable form in an ancillary file together with the arXiv submission of this article.

\bibliography{biblio}

\providecommand{\href}[2]{#2}\begingroup\raggedright\begin{thebibliography}{10}

\bibitem{Aad2012}
{The Atlas Collaboration}, {\it {Observation of a new particle in the search
  for the Standard Model Higgs boson with the ATLAS detector at the LHC}},
  {\em Physics Letters, Section B: Nuclear, Elementary Particle and High-Energy
  Physics} {\bf 716} (2012), no.~1 1--29,
  [\href{http://xxx.lanl.gov/abs/1207.7214}{{\tt 1207.7214}}].

\bibitem{Chatrchyan2012}
{The CMS Collabortaion}, {\it {Observation of a new boson at a mass of 125 GeV
  with the CMS experiment at the LHC}},  {\em Phys.Lett.} {\bf B716} (2012)
  30--61.

\bibitem{Dawson:1990zj}
S.~Dawson, {\it {Radiative corrections to Higgs boson production}},  {\em Nucl.
  Phys.} {\bf B359} (1991) 283--300.

\bibitem{Spira:1995rr}
M.~Spira, A.~Djouadi, D.~Graudenz, and P.~M. Zerwas, {\it {Higgs boson
  production at the LHC}},  {\em Nucl. Phys.} {\bf B453} (1995) 17--82,
  [\href{http://xxx.lanl.gov/abs/hep-ph/9504378}{{\tt hep-ph/9504378}}].

\bibitem{Wilczek1977}
F.~Wilczek, {\it {Decays of Heavy Vector Mesons into Higgs Particles}},  {\em
  Physical Review Letters} {\bf 39} (1977), no.~21 1304--1306.

\bibitem{Harlander:2002wh}
R.~V. Harlander and W.~B. Kilgore, {\it {Next-to-next-to-leading order Higgs
  production at hadron colliders}},  {\em Phys. Rev. Lett.} {\bf 88} (2002)
  201801, [\href{http://xxx.lanl.gov/abs/hep-ph/0201206}{{\tt
  hep-ph/0201206}}].

\bibitem{Anastasiou2002}
C.~Anastasiou and K.~Melnikov, {\it {Higgs boson production at hadron colliders
  in NNLO QCD}},  {\em Nuclear Physics B} {\bf 646} (dec, 2002) 220--256,
  [\href{http://xxx.lanl.gov/abs/0207004}{{\tt 0207004}}].

\bibitem{Ravindran:2003um}
V.~Ravindran, J.~Smith, and W.~L. van Neerven, {\it {NNLO corrections to the
  total cross-section for Higgs boson production in hadron hadron collisions}},
   {\em Nucl. Phys.} {\bf B665} (2003) 325--366,
  [\href{http://xxx.lanl.gov/abs/hep-ph/0302135}{{\tt hep-ph/0302135}}].

\bibitem{Anastasiou:2014vaa}
C.~Anastasiou, C.~Duhr, F.~Dulat, E.~Furlan, T.~Gehrmann, F.~Herzog, and
  B.~Mistlberger, {\it {Higgs boson gluon--fusion production at threshold in
  N$^3$LO QCD}},  {\em Phys. Lett.} {\bf B737} (2014) 325--328,
  [\href{http://xxx.lanl.gov/abs/1403.4616}{{\tt 1403.4616}}].

\bibitem{Anastasiou:2014lda}
C.~Anastasiou, C.~Duhr, F.~Dulat, E.~Furlan, T.~Gehrmann, F.~Herzog, and
  B.~Mistlberger, {\it {Higgs boson gluon-fusion production beyond threshold in
  N$^{3}$LO QCD}},  {\em JHEP} {\bf 03} (2015) 091,
  [\href{http://xxx.lanl.gov/abs/1411.3584}{{\tt 1411.3584}}].

\bibitem{Anastasiou:2015ema}
C.~Anastasiou, C.~Duhr, F.~Dulat, F.~Herzog, and B.~Mistlberger, {\it {Higgs
  Boson Gluon-Fusion Production in QCD at Three Loops}},  {\em Phys. Rev.
  Lett.} {\bf 114} (2015) 212001,
  [\href{http://xxx.lanl.gov/abs/1503.06056}{{\tt 1503.06056}}].

\bibitem{Anastasiou:2016cez}
C.~Anastasiou, C.~Duhr, F.~Dulat, E.~Furlan, T.~Gehrmann, F.~Herzog,
  A.~Lazopoulos, and B.~Mistlberger, {\it {High precision determination of the
  gluon fusion Higgs boson cross-section at the LHC}},  {\em JHEP} {\bf 05}
  (2016) 058, [\href{http://xxx.lanl.gov/abs/1602.00695}{{\tt 1602.00695}}].

\bibitem{deFlorian:2016spz}
{\bf LHC Higgs Cross Section Working Group} Collaboration, D.~de~Florian {\em
  et~al.}, {\it {Handbook of LHC Higgs Cross Sections: 4. Deciphering the
  Nature of the Higgs Sector}},  \href{http://xxx.lanl.gov/abs/1610.07922}{{\tt
  1610.07922}}.

\bibitem{Harlander:2016hcx}
R.~V. Harlander, S.~Liebler, and H.~Mantler, {\it {SusHi Bento: Beyond NNLO and
  the heavy-top limit}},  {\em Comput. Phys. Commun.} {\bf 212} (2017)
  239--257, [\href{http://xxx.lanl.gov/abs/1605.03190}{{\tt 1605.03190}}].

\bibitem{Chen2015}
X.~Chen, T.~Gehrmann, E.~Glover, and M.~Jaquier, {\it {Precise QCD predictions
  for the production of Higgs + jet final states}},  {\em Physics Letters B}
  {\bf 740} (jan, 2015) 147--150.

\bibitem{Boughezal2015b}
R.~Boughezal, F.~Caola, K.~Melnikov, F.~Petriello, and M.~Schulze, {\it {Higgs
  boson production in association with a jet at next-to-next-to-leading
  order}},  {\em Physical Review Letters} {\bf 115} (apr, 2015) 082003,
  [\href{http://xxx.lanl.gov/abs/1504.07922}{{\tt 1504.07922}}].

\bibitem{Boughezal2015a}
R.~Boughezal, C.~Focke, W.~Giele, X.~Liu, and F.~Petriello, {\it {Higgs boson
  production in association with a jet using jettiness subtraction}},  {\em
  Physics Letters B} {\bf 748} (sep, 2015) 5--8.

\bibitem{Banfi2016}
A.~Banfi, F.~Caola, F.~A. Dreyer, P.~F. Monni, G.~P. Salam, G.~Zanderighi, and
  F.~Dulat, {\it Jet-vetoed higgs cross section in gluon fusion at n3lo+nnll
  with small-r resummation},  {\em Journal of High Energy Physics} {\bf 2016}
  (2016), no.~4 49.

\bibitem{Anastasiou:2005cb}
C.~Anastasiou and A.~Daleo, {\it {Numerical evaluation of loop integrals}},
  {\em JHEP} {\bf 10} (2006) 031,
  [\href{http://xxx.lanl.gov/abs/hep-ph/0511176}{{\tt hep-ph/0511176}}].

\bibitem{Binoth:2000ps}
T.~Binoth and G.~Heinrich, {\it {An automatized algorithm to compute infrared
  divergent multiloop integrals}},  {\em Nucl. Phys.} {\bf B585} (2000)
  741--759, [\href{http://xxx.lanl.gov/abs/hep-ph/0004013}{{\tt
  hep-ph/0004013}}].

\bibitem{Hepp:1966eg}
K.~Hepp, {\it {Proof of the Bogolyubov-Parasiuk theorem on renormalization}},
  {\em Commun. Math. Phys.} {\bf 2} (1966) 301--326.

\bibitem{Roth:1996pd}
M.~Roth and A.~Denner, {\it {High-energy approximation of one loop Feynman
  integrals}},  {\em Nucl. Phys.} {\bf B479} (1996) 495--514,
  [\href{http://xxx.lanl.gov/abs/hep-ph/9605420}{{\tt hep-ph/9605420}}].

\bibitem{Boughezal2011}
R.~Boughezal, K.~Melnikov, and F.~Petriello, {\it {A subtraction scheme for
  NNLO computations}},  {\em Physical Review D} {\bf 85} (nov, 2011) 034025,
  [\href{http://xxx.lanl.gov/abs/1111.7041}{{\tt 1111.7041}}].

\bibitem{Boughezal2015}
R.~Boughezal, X.~Liu, and F.~Petriello, {\it {The N-jettiness soft function at
  next-to-next-to-leading order}},  {\em Physical Review D} {\bf 91} (apr,
  2015) 094035, [\href{http://xxx.lanl.gov/abs/1504.02540}{{\tt 1504.02540}}].

\bibitem{Catani2007}
S.~Catani and M.~Grazzini, {\it {Next-to-Next-to-Leading-Order Subtraction
  Formalism in Hadron Collisions and its Application to Higgs-Boson Production
  at the Large Hadron Collider}},  {\em Physical Review Letters} {\bf 98} (may,
  2007) 222002, [\href{http://xxx.lanl.gov/abs/0703012}{{\tt 0703012}}].

\bibitem{Gaunt2015}
J.~R. Gaunt, M.~Stahlhofen, F.~J. Tackmann, and J.~R. Walsh, {\it {N-jettiness
  subtractions for NNLO QCD calculations}},  {\em Journal of High Energy
  Physics} {\bf 2015} (sep, 2015) 58.

\bibitem{Baernreuther2012}
P.~Baernreuther, M.~Czakon, and A.~Mitov, {\it {Percent level precision physics
  at the Tevatron: first genuine NNLO QCD corrections to q qbar -{\textgreater}
  t tbar + X}},  {\em Physical Review Letters} {\bf 109} (apr, 2012) 132001,
  [\href{http://xxx.lanl.gov/abs/1204.5201}{{\tt 1204.5201}}].

\bibitem{Caola2017}
F.~Caola, K.~Melnikov, and R.~R{\"{o}}ntsch, {\it {Nested soft-collinear
  subtractions in NNLO QCD computations}},
  \href{http://xxx.lanl.gov/abs/1702.01352}{{\tt 1702.01352}}.

\bibitem{i2007}
G.~Somogyi, Z.~Tr{\'{o}}cs{\'{a}}nyi, and V.~D. Duca, {\it {A subtraction
  scheme for computing QCD jet cross sections at NNLO: regularization of
  doubly-real emissions}},  {\em Journal of High Energy Physics} {\bf 2007}
  (jan, 2007) 070--070.

\bibitem{Ridder2005}
A.~G.-D. Ridder, T.~Gehrmann, and E.~N. Glover, {\it {Antenna subtraction at
  NNLO}},  {\em Journal of High Energy Physics} {\bf 2005} (sep, 2005)
  056--056.

\bibitem{Cacciari:2015jma}
M.~Cacciari, F.~A. Dreyer, A.~Karlberg, G.~P. Salam, and G.~Zanderighi, {\it
  {Fully Differential Vector-Boson-Fusion Higgs Production at
  Next-to-Next-to-Leading Order}},  {\em Phys. Rev. Lett.} {\bf 115} (2015),
  no.~8 082002, [\href{http://xxx.lanl.gov/abs/1506.02660}{{\tt 1506.02660}}].

\bibitem{Anastasiou2010}
C.~Anastasiou, F.~Herzog, and A.~Lazopoulos, {\it {On the factorization of
  overlapping singularities at NNLO}},  {\em Journal of High Energy Physics}
  {\bf 2011} (nov, 2010) 38, [\href{http://xxx.lanl.gov/abs/1011.4867}{{\tt
  1011.4867}}].

\bibitem{Dulat:2017aa}
F.~Dulat, S.~Lionetti, B.~Mistlberger, A.~Pelloni, and C.~Specchia, {\it
  {Higgs-differential cross section at NNLO in dimensional regularisation}},
  {\em JHEP} {\bf 07} (2017) 017,
  [\href{http://xxx.lanl.gov/abs/1704.08220}{{\tt 1704.08220}}].

\bibitem{Anastasiou:2002qz}
C.~Anastasiou, L.~Dixon, and K.~Melnikov, {\it {NLO Higgs boson rapidity
  distributions at hadron colliders}},  {\em Nuclear Physics B - Proceedings
  Supplements} {\bf 116} (mar, 2003) 193--197.

\bibitem{Anastasiou:2003yy}
C.~Anastasiou, L.~Dixon, K.~Melnikov, and F.~Petriello, {\it {Dilepton Rapidity
  Distribution in the Drell-Yan Process at Next-to-Next-to-Leading Order in
  QCD}},  {\em Physical Review Letters} {\bf 91} (oct, 2003) 182002.

\bibitem{Ravindran:2006bu}
V.~Ravindran, J.~Smith, and W.~L. van Neerven, {\it {QCD threshold corrections
  to di-lepton and Higgs rapidity distributions beyond $N^{2}$ LO}},  {\em
  Nucl. Phys.} {\bf B767} (2007) 100--129,
  [\href{http://xxx.lanl.gov/abs/hep-ph/0608308}{{\tt hep-ph/0608308}}].

\bibitem{Ahmed:2014uya}
T.~Ahmed, M.~K. Mandal, N.~Rana, and V.~Ravindran, {\it {Rapidity Distributions
  in Drell-Yan and Higgs Productions at Threshold to Third Order in QCD}},
  {\em Phys. Rev. Lett.} {\bf 113} (2014) 212003,
  [\href{http://xxx.lanl.gov/abs/1404.6504}{{\tt 1404.6504}}].

\bibitem{Banerjee:2017cfc}
P.~Banerjee, G.~Das, P.~K. Dhani, and V.~Ravindran, {\it {Threshold resummation
  of rapidity distribution in Higgs production at NNLO+NNLL}},
  \href{http://xxx.lanl.gov/abs/1708.05706}{{\tt 1708.05706}}.

\bibitem{Li:2016axz}
Y.~Li, D.~Neill, and H.~X. Zhu, {\it {An Exponential Regulator for Rapidity
  Divergences}},  {\em Submitted to: Phys. Rev. D} (2016)
  [\href{http://xxx.lanl.gov/abs/1604.00392}{{\tt 1604.00392}}].

\bibitem{Li:2016ctv}
Y.~Li and H.~X. Zhu, {\it {Bootstrapping Rapidity Anomalous Dimensions for
  Transverse-Momentum Resummation}},  {\em Phys. Rev. Lett.} {\bf 118} (2017),
  no.~2 022004, [\href{http://xxx.lanl.gov/abs/1604.01404}{{\tt 1604.01404}}].

\bibitem{Anastasiou:2013srw}
C.~Anastasiou, C.~Duhr, F.~Dulat, and B.~Mistlberger, {\it {Soft triple-real
  radiation for Higgs production at N3LO}},
  \href{http://xxx.lanl.gov/abs/1302.4379}{{\tt 1302.4379}}.

\bibitem{Anastasiou:2013mca}
C.~Anastasiou, C.~Duhr, F.~Dulat, F.~Herzog, and B.~Mistlberger, {\it
  {Real-virtual contributions to the inclusive Higgs cross-section at
  $N^3LO$}},  {\em JHEP} {\bf 12} (2013) 088,
  [\href{http://xxx.lanl.gov/abs/1311.1425}{{\tt 1311.1425}}].

\bibitem{Anastasiou:2015yha}
C.~Anastasiou, C.~Duhr, F.~Dulat, E.~Furlan, F.~Herzog, and B.~Mistlberger,
  {\it {Soft expansion of double-real-virtual corrections to Higgs production
  at N$^{3}$LO}},  {\em JHEP} {\bf 08} (2015) 051,
  [\href{http://xxx.lanl.gov/abs/1505.04110}{{\tt 1505.04110}}].

\bibitem{Chetyrkin:1997un}
K.~G. Chetyrkin, B.~A. Kniehl, and M.~Steinhauser, {\it {Decoupling relations
  to O (alpha-s**3) and their connection to low-energy theorems}},  {\em Nucl.
  Phys.} {\bf B510} (1998) 61--87,
  [\href{http://xxx.lanl.gov/abs/hep-ph/9708255}{{\tt hep-ph/9708255}}].

\bibitem{Schroder:2005hy}
Y.~Schroder and M.~Steinhauser, {\it {Four-loop decoupling relations for the
  strong coupling}},  {\em JHEP} {\bf 01} (2006) 051,
  [\href{http://xxx.lanl.gov/abs/hep-ph/0512058}{{\tt hep-ph/0512058}}].

\bibitem{Chetyrkin:2005ia}
K.~Chetyrkin, J.~K{\"{u}}hn, and C.~Sturm, {\it {QCD decoupling at four
  loops}},  {\em Nuclear Physics B} {\bf 744} (jun, 2006) 121--135.

\bibitem{Kramer:1996iq}
M.~Kramer, E.~Laenen, and M.~Spira, {\it {Soft gluon radiation in Higgs boson
  production at the LHC}},  {\em Nucl. Phys.} {\bf B511} (1998) 523--549,
  [\href{http://xxx.lanl.gov/abs/hep-ph/9611272}{{\tt hep-ph/9611272}}].

\bibitem{Graudenz:1992pv}
D.~Graudenz, M.~Spira, and P.~M. Zerwas, {\it {QCD corrections to Higgs boson
  production at proton proton colliders}},  {\em Phys. Rev. Lett.} {\bf 70}
  (1993) 1372--1375.

\bibitem{Gehrmann2010}
T.~Gehrmann, E.~W.~N. Glover, T.~Huber, N.~Ikizlerli, and C.~Studerus, {\it
  {Calculation of the quark and gluon form factors to three loops in QCD}},
  {\em Journal of High Energy Physics} {\bf 2010} (jun, 2010) 94.

\bibitem{Duhr:2013msa}
C.~Duhr and T.~Gehrmann, {\it {The two-loop soft current in dimensional
  regularization}},  {\em Phys. Lett.} {\bf B727} (2013) 452--455,
  [\href{http://xxx.lanl.gov/abs/1309.4393}{{\tt 1309.4393}}].

\bibitem{Duhr:2014nda}
C.~Duhr, T.~Gehrmann, and M.~Jaquier, {\it {Two-loop splitting amplitudes and
  the single-real contribution to inclusive Higgs production at N$^3$LO}},
  {\em JHEP} {\bf 02} (2015) 077,
  [\href{http://xxx.lanl.gov/abs/1411.3587}{{\tt 1411.3587}}].

\bibitem{Nogueira1993}
P.~Nogueira, {\it {Automatic Feynman Graph Generation}},  {\em Journal of
  Computational Physics} {\bf 105} (apr, 1993) 279--289.

\bibitem{Bauer2000}
C.~W. Bauer, R.~Kreckel, and A.~Frink, {\it {Introduction to the GiNaC
  framework for symbolic computation within the C++ programming language}},
  {\em J.Symb.Comput.} {\bf 33} (2000) 1.

\bibitem{Li:2014bfa}
Y.~Li, A.~von Manteuffel, R.~M. Schabinger, and H.~X. Zhu, {\it {N$^3$LO Higgs
  boson and Drell-Yan production at threshold: The one-loop two-emission
  contribution}},  {\em Phys. Rev.} {\bf D90} (2014), no.~5 053006,
  [\href{http://xxx.lanl.gov/abs/1404.5839}{{\tt 1404.5839}}].

\bibitem{Anastasiou2003}
C.~Anastasiou and K.~Melnikov, {\it {Pseudoscalar Higgs boson production at
  hadron colliders in next-to-next-to-leading order QCD}},  {\em Physical
  Review D} {\bf 67} (feb, 2003) 037501.

\bibitem{Anastasiou2004a}
C.~Anastasiou, L.~Dixon, K.~Melnikov, and F.~Petriello, {\it {High-precision
  QCD at hadron colliders: Electroweak gauge boson rapidity distributions at
  next-to-next-to leading order}},  {\em Physical Review D} {\bf 69} (may,
  2004) 094008.

\bibitem{Kotikov1991}
A.~Kotikov, {\it {Differential equations method. New technique for massive
  Feynman diagram calculation}},  {\em Physics Letters B} {\bf 254} (jan, 1991)
  158--164.

\bibitem{Gehrmann2000}
T.~Gehrmann and E.~Remiddi, {\it {Differential equations for two-loop
  four-point functions}},  {\em Nuclear Physics B} {\bf 580} (jul, 2000)
  485--518.

\bibitem{Henn2013}
J.~M. Henn, {\it {Multiloop Integrals in Dimensional Regularization Made
  Simple}},  {\em Physical Review Letters} {\bf 110} (jun, 2013) 251601.

\bibitem{REMIDDI2000}
E.~Remiddi and J.~A.~M. Vermaseren, {\it {Harmonic Polylogarithms}},  {\em
  International Journal of Modern Physics A} {\bf 15} (feb, 2000) 725--754.

\bibitem{Beneke:1997zp}
M.~Beneke and V.~A. Smirnov, {\it {Asymptotic expansion of Feynman integrals
  near threshold}},  {\em Nucl. Phys.} {\bf B522} (1998) 321--344,
  [\href{http://xxx.lanl.gov/abs/hep-ph/9711391}{{\tt hep-ph/9711391}}].

\bibitem{Harland-Lang:2014zoa}
L.~A. Harland-Lang, A.~D. Martin, P.~Motylinski, and R.~S. Thorne, {\it {Parton
  distributions in the LHC era: MMHT 2014 PDFs}},  {\em Eur. Phys. J.} {\bf
  C75} (2015), no.~5 204, [\href{http://xxx.lanl.gov/abs/1412.3989}{{\tt
  1412.3989}}].

\bibitem{Herzog:2014wja}
F.~Herzog and B.~Mistlberger, {\it {The Soft-Virtual Higgs Cross-section at
  N3LO and the Convergence of the Threshold Expansion}},  in {\em {Proceedings,
  49th Rencontres de Moriond on QCD and High Energy Interactions: La Thuile,
  Italy, March 22-29, 2014}}, pp.~57--60, 2014.
\newblock \href{http://xxx.lanl.gov/abs/1405.5685}{{\tt 1405.5685}}.

\bibitem{Dulat:2015mca}
S.~Dulat, T.-J. Hou, J.~Gao, M.~Guzzi, J.~Huston, P.~Nadolsky, J.~Pumplin,
  C.~Schmidt, D.~Stump, and C.~P. Yuan, {\it {New parton distribution functions
  from a global analysis of quantum chromodynamics}},  {\em Phys. Rev.} {\bf
  D93} (2016), no.~3 033006, [\href{http://xxx.lanl.gov/abs/1506.07443}{{\tt
  1506.07443}}].

\bibitem{Ball:2014uwa}
{\bf NNPDF} Collaboration, R.~D. Ball {\em et~al.}, {\it {Parton distributions
  for the LHC Run II}},  {\em JHEP} {\bf 04} (2015) 040,
  [\href{http://xxx.lanl.gov/abs/1410.8849}{{\tt 1410.8849}}].

\bibitem{Alekhin:2017kpj}
S.~Alekhin, J.~Blümlein, S.~Moch, and R.~Placakyte, {\it {Parton distribution
  functions, $\alpha_s$, and heavy-quark masses for LHC Run II}},  {\em Phys.
  Rev.} {\bf D96} (2017), no.~1 014011,
  [\href{http://xxx.lanl.gov/abs/1701.05838}{{\tt 1701.05838}}].

\bibitem{Butterworth:2015oua}
J.~Butterworth {\em et~al.}, {\it {PDF4LHC recommendations for LHC Run II}},
  {\em J. Phys.} {\bf G43} (2016) 023001,
  [\href{http://xxx.lanl.gov/abs/1510.03865}{{\tt 1510.03865}}].

\bibitem{Buckley:2014ana}
A.~Buckley, J.~Ferrando, S.~Lloyd, K.~Nordström, B.~Page, M.~Rüfenacht,
  M.~Schönherr, and G.~Watt, {\it {LHAPDF6: parton density access in the LHC
  precision era}},  {\em Eur. Phys. J.} {\bf C75} (2015) 132,
  [\href{http://xxx.lanl.gov/abs/1412.7420}{{\tt 1412.7420}}].

\end{thebibliography}\endgroup
\bibliographystyle{JHEP}

\end{document}